\documentclass[11pt, a4paper,twoside]{article}
\usepackage{chngcntr}
\counterwithin{figure}{section}
\counterwithin{table}{section}

\usepackage{lscape}

\usepackage{amsmath}
\usepackage{amssymb} 
\usepackage{graphicx}
\usepackage{svg}

\usepackage[margin=1.4in]{geometry}
\usepackage{chicago}
\usepackage{array,color,colortbl,xspace}
\usepackage{url}

 \usepackage[section]{placeins}
 \makeatletter
 \AtBeginDocument{%
   \expandafter\renewcommand\expandafter\subsection\expandafter{%
     \expandafter\@fb@secFB\subsection
   }%
 }
 \makeatother

\DeclareMathOperator*{\argmin}{argmin}

\def\thec{f}\def\thec{c}
\def\thep{q}\def\thep{p}\def\thep{2}
\def\convolutional{factorial\xspace}\def\convolutional{convolutional\xspace}

\def\functional{interpolation\xspace}\def\functional{functional\xspace}
\def\Functional{Interpolation\xspace}\def\Functional{Functional\xspace}

\def\b{}\def\b{\textcolor{blue}}

\def\point{|point|}\def\point{m}

\def\fact{|fact|}\def\fact{f}

 \def\available{non-missing\xspace}\def\available{available\xspace}

\begin{document}

\title{Nowcasting Networks\textsuperscript{\dag}}
\author{Marc Chataigner\textsuperscript{1}, St\'ephane Cr\'epey\textsuperscript{2}, and Jiang Pu\textsuperscript{3}} 

{\let\thefootnote\relax\footnotetext{\textsuperscript{1} \textit{LaMME, Universit\'e de Paris-Saclay}}}

{\let\thefootnote\relax\footnotetext{\textsuperscript{2} \textit{LPSM, Universit\'e de Paris}}}

{\let\thefootnote\relax\footnotetext{\textsuperscript{3} \textit{Léonard de Vinci Pôle Universitaire, Research Center, Paris La Défense}}}

{\let\thefootnote\relax\footnotetext{\textsuperscript{\dag} \textit{This article has been accepted for publication in Journal of Computational Finance, published by
Incisive Media Ltd.}}}
{\let\thefootnote\relax\footnotetext{{{\it Acknowledgement:}  
This research has been conducted with the support of the Research Initiative “Mod\'elisation des march\'es actions, obligations
et d\'eriv\'es”
 financed by HSBC France under the aegis of the Europlace Institute of Finance. All data sets in our case studies have been put at our disposal by HSBC. The views and opinions expressed in this presentation are those of the author
alone and do not necessarily reflect the views or policies of HSBC Investment
Bank, its subsidiaries or affiliates.
The research of Marc Chataigner is co-supported by a public grant as part of investissement d'avenir project, reference ANR-11-LABX-0056-LLH LabEx LMH. The authors would like
to thank  
Daniel Girard, Nicolas Grandchamp des Raux, Hugo Lebrun, and Guillaume Macey,
for their advice and encouragement during the conduct of this study. The authors also thank Maud Thomas for her bibliographical assistance regarding the outlier detection literature review of Section \ref{ss:outl-detn}.}}}

\date{\today}  
 
\maketitle

\begin{abstract}  
We devise a 
neural network based compression/completion methodology 
for financial nowcasting. The latter is meant in a broad sense encompassing completion of gridded values, 
interpolation, or outlier detection, in the context of financial time series of curves or surfaces (also applicable in higher dimensions, at least in theory). 
In particular, we introduce an original architecture amenable to the treatment of  
data defined at variable grid nodes (by far the most common situation in financial nowcasting applications, so that  PCA or classical autoencoder methods are not applicable).
This is illustrated by three case studies on real data sets. 
First, we introduce our approach  
on repo curves data (with moving time-to-maturity as calendar time passes).
Second, we show that our approach outperforms  elementary interpolation
benchmarks on an equity derivative surfaces data set (with moving time-to-maturity again). 
We also obtain a satisfying performance for outlier detection and surface completion. 
Third, we benchmark our approach against PCA on
at-the-money swaption surfaces  redefined at constant expiry/tenor grid nodes.
Our approach is then shown to perform as well as (even if not obviously better than)
the PCA (which, however, is not be applicable to the native, raw data defined on a moving time-to-expiry grid).

\end{abstract}

\def\keywordname{{\bfseries Keywords:}}
\def\keywords#1{\par\addvspace\baselineskip\noindent\keywordname\enspace
\ignorespaces#1}
 
\vspace{2mm}
 

\newpage 



\section{Introduction}\label{s:intro}

In this paper, we devise 
a neural network based methodology
for financial nowcasting. The latter is meant in a broad sense encompassing completion of gridded values,
interpolation, or outlier detection, in the context of financial time series of curves or surfaces. 
Toward this end we develop
a generic two-step methodology, whereby a pre-processing compression stage is followed by a completion stage. Moreover, we detail two variations along this baseline, corresponding to two 
slightly
different perspectives and 
significantly distinct neural network
architectures.

As such our  approach is not bound to vectors and matrices. For generality and notational convenience it is
presented in the methodological part on arbitrary tensors
(but we do not address the strong aspect of dimension reduction that typically comes with genuine tensors as opposed to matrices and vectors).

Under the so called 
 \convolutional approach,
which is of the autoencoder type,
we assume that the information contained in an observed tensor can be encoded into a reduced set of 
variables, dubbed factors.
Conversely, given the factors, we can reconstruct the whole tensor with a decoder. 
As a limiting case, we obtain a linear, principal component  analysis (PCA) kind of approach,
%
but one itself implemented in the optimization training mode, as an autoencoder with linear activation functions (as opposed to spectral decomposition in the classical PCA case).

Under the so called
\functional approach, 
factors are rather used as a way to adjust a map taking as input a location (coordinates that may be part or not of the original tensor nodes) and returning the corresponding
reconstructed value.

The \convolutional approach is more particularly dedicated to completion of values on a fixed grid of coordinates,
whereas the \functional approach can handle moving grids, which corresponds to the vast majority of applications in financial nowcasting applications (unless the data have been transformed in a preprocessing stage to make them fit a fixed grid, entailing an undesirable layer of approximation). 
Moreover, in the \functional approach, including additional variables is straightfoward. 

The use of autoencoders as a nonlinear extension of the PCA can be traced back to the 1980s (see Chapter 14 in \citeN{DeepLearning} for a survey of autoencoder-based learning).
Autoencoders  have also
already been used in data completion (see \citeN{kiran2018overview},  
\citeN{strub2016hybrid}). In contrast, the neural network architecture of our \functional approach is new to the best of our knowledge. 

%

At the intersection between neural networks and finance, the related paper by
\citeN{kondratyev2018learning} 
is more about forecasting. 
Accordingly, we work in a mostly unsupervised setting, whereas \citeN{kondratyev2018learning} is in a mostly supervised setting. \citeN{kondratyev2018learning} predicts a new curve given a shock on a curve. The neural network is trained for shocks applied to a particular location.
Hence, to consider a new shock, the model needs to be retrained.
In contrast, our convolutional network has a latent structure capturing interdependencies between all points in the grid.
This is even more obvious in the case of our \functional approach, where extra variables can be provided as direct inputs to the model.

Autoencoders (hence, unsupervised learning)  are also considered in Section 5.4 in \citeN{kondratyev2018learning}. However, this is then with a focus on curve regularization on a fixed grid, 
which can be done directly by decoding. The completion problem that we are dealing in this work is more general and it requires one additional layer of numerical optimization.
Moreover, \citeN{kondratyev2018learning} only deals with the univariate case of curves, for which spatial regularity is a much less challenging issue.


The paper is outlined as follows.
Sections \ref{s:problems} and \ref{s:models} introduce the problems and models. By the latter, we mean different algorithmic strategies and neural network architectures that can be used for addressing the former. 
Section \ref{s:exp} lays
an experimental setup putting the different models on comparable grounds.   
Sections \ref{s:repo}, \ref{s:eq} and \ref{s:swn} present repo curves, equity derivative implied volatility surfaces and
at-the-money swaption implied volatility surfaces case studies on real data sets.
Section \ref{ss:concl} concludes and discusses
further research perspectives in connection with the quantitative finance and machine learning literatures.

Any notation of the form $\min_{x}\Lambda(x,y)$ means 
that we minimize in $x$ a loss $\Lambda$ given the value $y$ of additional parameters; $x^\star$ then refers to a numerical minimizer of $\Lambda(x,y)$ (which is typically nonconvex in $x$), for this given $y$.

\section{Problems}\label{s:problems}

We consider a data set consisting of a time series of observations, each consisting of $\point$ points, or
 features,
structured as a multivariate tensor.
By the latter, we mean a discretized 
cube  (curve or surface in our case studies, but the methodology is generic) of values of homogenous quantities, such as 
rates of different terms, implied volatilities of different strikes and maturities, etc., defined at each tensor grid node. 



\subsection{Compression}\label{ss:compr}

The compression problem is mainly a pre-processing stage that aims at reducing the dimensionality $\point$ of a feature space, i.e. the number of grid nodes in each tensor (here assumed constant across observations $\omega$, see Section \ref{ss:functio} regarding the variant of the functional approach with a possibly variable $m_\omega$).  
Assume that each observation takes its values in (a subset of) ${\mathbb{R}}^{\point}$. 
We call encoder $E$ any injective map from a relevant subset  $\mathcal{S}$ of ${\mathbb{R}}^{\point}$ to a space 
$\mathbb{R}^{\fact}$
of factors,  
where $\fact \ll \point$ is the number of factors.
Conversely, one would like to  be able to reconstruct the $\point$ values of a tensor from any set of factors, or code, thanks to a 
map, called decoder, $D: \mathbb{R}^{\fact} \rightarrow 
\mathcal{S}$.
The compression challenge is to build $D$ and $E$ such that $D \circ E: \mathcal{S} \rightarrow \mathcal{S}$ is bijective and ``as close as possible to identity'' (cf.~\citeN[Chapter 14]{DeepLearning}).

The inspection of common financial time series of tensors suggests that, in their case, this challenge is somehow not unreasonable. Indeed, 
structural constraints often exist between the values at different  tensor nodes, e.g.~arbitrage pricing relationships throughout the option chain.
Moreover, usual financial tensors exhibit some spatial regularity, in the sense that values at grid nodes vary smoothly with respect to node location (think of interest rates with respect to their term or implied volatilities with respect to the maturity and strike of an option).
In addition, some coordinates may have a regularizing effect. For instance, in the region of large expiries, the at-the-money swaption implied volatility surface is mostly affected by translation moves (and not so much by steepening, etc.) as time passes  (see Section \ref{s:swn}).
Last, some (monotonicity, convexity,...) patterns are often apparent  (e.g. the well-known volatility smile in equity derivative, and some similar features in interest rate swaption implied volatility surfaces, cf.~Figure \ref{fig:sub1}).

Both maps $E$ and $D$ are sought within classes of neural networks with respective parameters $\varepsilon$ and $\delta$,
 collectively denoted by $\theta$.
The motivation for using neural networks in this context is their nonparametric (or, at least, 
 very expressive) and nonlinear features. Gaussian processes for instance would be much less flexible, with only a few, e.g. two, kernel hyperparameters for squared exponential kernel to calibrate a full data set of thousands of tensors.

We include into $\theta$ weights, biases, as well as any variable calibrated during the compression stage. Denoting $E=E_\varepsilon$ and $D=D_\delta$ in reference to this parameterization, the compression stage is the training of the neural networks according to the following optimization problem:
\begin{equation}\label{eq:Compression}
\min_{\theta=(\delta,\varepsilon)
}
\sum_{\omega\in\Omega}\sum_{(n,y)\in\omega}
\left(y-\Big(D_{\delta} \big(E_{\varepsilon}( \omega )\big)\Big)_n\right)^{\thep},
\end{equation}
where $\Omega$ stands for the training data set (cf.~Section \ref{s:exp}).

Certain additional properties are desirable for 
$D$ and $E$. 
The parameterization $\theta$ should allow for a robust and fast numerical solution to the problem \eqref{eq:Compression}. This may be harder to achieve
for some deep neural networks too sensitive to the initialization of their parameters. In particular, two similar tensors should give rise to similar codes and vice versa, i.e.~we want $D$ and $E$ to be ``sufficiently smooth'' in such way as to preserve distance in the subspace.   

\subsection{Completion}\label{ss:compl}


Having found a parametrization $\theta^\star=(\delta^\star,\varepsilon^\star)$ that ensures a satisfying reconstruction loss in \eqref{eq:Compression}, 
the completion task
consists in the exploitation of $D_{\delta^{\star}}$ in 
order to find the missing values of an 
incomplete observation $\omega$ (of the current day, say, to be completed based on the complete observations of the previous days, used as training set).

Toward this end, we
introduce the following optimization problem:
\begin{equation}\label{eq:Completion}
\min\limits_{c 
} 
\sum_{\underset{\b{\text{ }  } }{(n,y)  \in\omega \mbox{\footnotesize}} }
\Big(y-\big(D_{\delta }( \thec)\big)_n\Big)^{\thep} ,
\end{equation}
considered for $\delta=\delta^{\star}$.
The completed tensor is then defined as the image $D_{\delta^{\star}}(c^{\star})$ of the code $c^{\star}$ by the decoder $D_{\delta^{\star}}$. Obviously, the more missing values, the harder the completion task (higher overfitting risk, unless some appropriate regularization is used).

Note that, thanks to the compression step, the number of variables to estimate is drastically reduced in \eqref{eq:Completion}, to some reference number, i.e.~the dimensionality of $c$ (e.g.~4, 15, or 8 in our repo, equity index derivative and interest-rate swaption case studies), independent of the  number of unknowns in the native,
``uncompressed'' completion problem (such as the number of missing implied volatility values in a to-be-completed surface).
Moreover, a factorial representation with $\fact \ll \point$ filters out the unlikely tensors (as outlined by the reconstruction error from our neural networks,
cf.~Section \ref{ss:outl-detn}) that could otherwise arise from a decoding due to the ill-posedness of large-scale arg-minimization problems.
The regularity of the map $D_{\delta^{\star}}$ can sometimes be exploited to ease the completion, by initializing the numerical solution of \eqref{eq:Completion} with the encoding of the last fully observed (e.g. already completed) tensor.


\paragraph{Literature Review} The literature on completion primarily deals  with data structured on a fixed grid. This means that columns in the data set refer to the same feature  (in our case: financial instrument). This is not consistent with most financial nowcasting applications, for which, in particular, the time-to-maturity decreases with calendar time. To the best of our knowledge, only naive interpolation methods on a given tensor, without possible exploitation of a data set, are available in the case of a moving grid. 

The standard completion framework relies on a low rank representation of the data set (see \citeN{nguyen2019low}). Along this line (but on a possibly moving grid), we compress each observation in a code which can be seen as a latent vector. However, 
in contrast with methods such as SVD, alternating least squares (see \citeN{hastie2015matrix}), or denoising autoencoders (see \citeN{strub2015collaborative}), which learn a user matrix, we do not consider the interaction between  the observations (i.e. the dynamics): at this stage at least, we focus on the interaction between the variables (instruments).

Finally, standard completion methods in recommender systems assume missing completely at random (MCAR) values dispersed throughout the whole data set. In our case studies missing values are located completely at random but only for the current observation.
\subsection{Outlier Detection}\label{ss:outl-detn}

 \citeN{hawkins1980identification} defines outliers as ``observation which deviates so much from other observations as to arouse suspicion it was generated by a different mechanism". Outlier detection is of course a crucial issue in finance. For instance, investment banks receive market information from a data provider. Sometimes, the data can be polluted with errors of various sources, or ``different mechanism", whether it is
data feed bugs, fat finger of other market participants, or failure from computation processes (for instance, implicitation of volatility surface from option prices). It can be either a punctual outlier, i.e.~a single value of the tensor is too far away from what it should be, or the whole tensor may have a shape that is very unlikely.

To detect the punctual outliers, many simple methods are available, based on smoothness metrics or
on	
historical percentile ranges of the values. To detect shape outliers, some criteria can be checked for very specific data sets, e.g.~non-arbitrage butterfly/calendar spread conditions in the case of option prices.

Here we propose a general method to detect both punctual outliers and shape aberration. The functional variant  of our method works on an unstructured grid.

To say that the tensors generated from the ``normal mechanism" is of a certain form is equivalent to say that the mechanism generates values that lie in a sub-manifold $\mathcal{S}$ of the initial feature space (cf.~Section \ref{ss:compr}). Finding this sub-manifold is equivalent to detecting anomalies. From this point of view, anomaly detection and compression/decompression are two sides of the same coin. Indeed, from an information theory point of view, there is an equivalence between being an anomaly and being hard to reconstruct (large reconstruction error in a lossy data compression setup, low or even negative compression rate in a lossless data compression setup): See the seminal paper by \citeN{shannon1948mathematical} or Chapter 4 in \citeN{mackay2003information}. That is, a compression/decompression setup provides a natural anomaly detection tool. 

Specifically,  we identify an outlier  as an observation whose reconstruction error (cf. \eqref{eq:Compression}) is above a predefined threshold.

Some key practical questions in outlier detection are how a threshold for outlier detection should  be chosen or how one  can validate the method. 
In principle this can only be addressed by human expertise. An expert would gradually diminish the threshold until the newly detected ‘outliers’ are no longer considered such by the expert. The method is valid and performs well if the outlier detection in a validation set is consistent with the expert view (so that, in particular, the  threshold is stable through time and does not need to be reassessed too frequently).
 
However, our compression methodology also provides a validation tool for the quality of our outlier detection method. Namely, one can corrupt some of the data (manually or in an automated fashion) and check whether the outlier detection procedure identifies the corrupted data. 
 
Our approach also provides guidance to a human expert for anomaly correction. Currently experts only rely on naive heuristics, such as interpolation between different points of a surface,  who cannot automatically exploit the overall data set of surfaces. In the outlier detection validation framework of the previous paragraph, one can also check whether the correction that our approach provides is closer to the true data than to the corrupted ones.
\paragraph{Literature Review} Among many related references on outlier detection:
\begin{itemize}
 \item \citeN{patcha2007overview},   \citeN{chandola2009anomaly},  \citeN{omar2013machine}, or 
 \citeN{anandakrishnan2018anomaly} provide surveys, the last one specialized in  finance and the next-to-last one on machine learning techniques;
  \item
 \citeN{lakhina2010feature} use PCA, 
  \citeN{an2015variational} variational autoencoders,  
  \citeN{schlegl2019f} generative adversarial networks,
  \citeN{lakhina2010feature} and \citeN{cappozzo2020anomaly} semi-supervised learning.
   \citeN{chaloner1988bayesian} and \citeN{cansado2008unsupervised} resort to Bayesian methodologies; 
  \item 
   \citeN{ro2015outlier} is about high-dimensional data,
 \citeN{anandakrishnan2018anomaly} about high dimensional big data,
   \citeN{rocke1996identification} about multivariate data,
 \citeN{goix2017sparse} and  \citeN{goix2015anomaly} about detection of anomalies among  extremes.
 \end{itemize}

\section{Models}\label{s:models}

The main innovation of this work is the functional approach. The description of the PCA and linear projection methods  are mainly provided so that the reader can compare carefully both frameworks. The PCA and linear projection methods are also used for benchmarking purposes in the swaption case study of Section \ref{s:swn} for which both approaches are available. The base PCA method is of course standard. The linear projection variant of it is detailed in Section \ref{ss:linear}. The description is short because the method falls directly under the umbrella of sections \ref{ss:compr} and \ref{ss:compl} (as opposed to the functional approach of section \ref{ss:functio}, which requires a specific development).

\subsection{The Convolutional (Autoencoder) Approach}\label{ss:convaa}


Typical autoencoder architectures are composed of two successive feedforward neural networks $E$ and $D$,
the encoder and the decoder.
Both networks can be constituted of several layers, intermediated by nonlinear activation functions,
with an overall bottleneck structure (to enforce compression in the middle).

Convolutional layers have been introduced for image processing and, more generally, any data structure represented as a tensor.
These networks aim to model the interactions between close points (whereas dense layers bind any output unit to all input units).
Spatial regularity properties are handled by a convolutional structure of the neural network architectures, whereby the only (non-zero) connections are between units corresponding to adjacent (in a suitable sense) grid nodes (cf.~Figure \ref{fig:convEncoder}).
The network then also uses fewer parameters, which reduces the complexity of the corresponding compression problem. 
For implementation details such as  kernels and padding, we refer to Chapter 9 in \citeN{DeepLearning}.

\subsection{The Linear Projection Approach}\label{ss:linear}

It is well known that an autoencoder with linear activation functions and an $ L_2 $ reconstruction error is equivalent to a PCA (see  Chapter 14 in \citeN{DeepLearning}).
As a limiting case of
the above, we consider a linear, PCA kind of benchmark,
%
but one itself implemented as an autoencoder with linear activation functions (as opposed to spectral decomposition for classical PCA implementation).
With respect to classical PCA (which will also be included in our case studies), this approach involves an additional bias parameter.
Moreover, it allows benefiting from the implicit regularization provided by early stopping in the related training procedure (see Section \ref{s:exp}), as opposed to a regularization provided by truncation of the lowest eigenvalues in spectral decomposition based PCA implementation.

\subsection{The \Functional Approach}\label{ss:functio} 

We introduce a variant of the above, especially suited to interpolation purposes (without reference to a fixed grid of nodes). This approach relies on a parameterized function $D=D_\delta(c,n)$ of a code $c$ and a node location $n$, where the latter no longer needs belong to a pre-determined grid. Here $\delta$ corresponds to the parameters of the decoder $D$, whereas the approach does not entail
any encoder
(at least, not explicitly). 

The compression is written as (compare with \eqref{eq:Compression}, using a similar notation as well as $C=(C_{\omega})_{\omega\in\Omega}$)
\begin{equation}\label{eqComprModu}
\min_{\delta
, 
C
}
\sum_{\omega\in\Omega}\sum_{(n,y)\in\omega}
\Big(y-D_{\delta}\big( C_{\omega},n\big)\Big)^{\thep} .
\end{equation}

Then, given a single, possibly partial observation $\omega$, the completion is given as (similar to \eqref{eq:Completion})
\begin{equation}\label{eqCompletion1}
\min\limits_{c 
} 
\sum_{(n,y)\in\omega}
\Big(y- D_{\delta }\big( \thec,n\big) \Big)^{\thep} ,
\end{equation}
considered for $\omega=\omega^\star$ and $\delta=\delta^\star$.
Importantly, for each given $\delta$, the minimization \eqref{eqComprModu} decouples into one (full observation) minimization \eqref{eqCompletion1} for each $\omega\in\Omega$.
Hence, the larger compression problem \eqref{eqComprModu} can be solved numerically as a succession of smaller problems \eqref{eqCompletion1}, in conjunction with gradient iterations in the direction of $\delta$. This ensures the scalability of the approach. It also makes it amenable to online learning.
The above observation also shows the consistency between \eqref{eqComprModu} and \eqref{eqCompletion1} in the sense that, if a full observation $\omega$ is used in \eqref{eqCompletion1}, it should yield $c^\star=C^\star_{\omega}$ (assuming global and unique minima to all problems for the sake of the argument).
 
Under this approach, dubbed \functional, 
 the decoder takes as input the location $ n $ of the point, in addition to the factors $ c $ (see Figure \ref{fig:jiangNN}).
It rebuilds each point individually, as per $ n \rightarrow D_{\delta} (c, n) $.
The network is thus able to interpolate between the nodes of the data grid.
The concept of neighborhood intervenes through the argument $ n $ of $ D $, but the parameterization $ \delta $ as well as the code $ c $ are common to all locations $n$. 
The compression \eqref{eqComprModu} can also accommodate incomplete data or discretization changes, i.e.~varying grids in the training data.
This feature allows training the \functional network with ``missing completely at random data'' 
 (MCAR, in the statistical missing data terminology).  

By comparison, under the 
convolutional approach of Section \ref{ss:convaa}, the concept of neighborhood intervenes through $ \theta=(\delta,\varepsilon) $, since each point of the grid is only sensitive to a subset of connections (the convolutional architecture only connects neighbouring points, cf. Figure  \ref{fig:convEncoder}).
 The encoding $ c $ is obtained directly thanks to $ E $, when the observation is complete, or by numerical completion (as always under the \functional approach) otherwise. 
 
 \subsection{Synthesis}
 
 To conclude this section,
 Tables \ref{tab:Summary} and \ref{tab:Summary2} summarize and put into perspective the different approaches referred to in the above.

\begin{table}[!htbp]
\centering
\resizebox{0.6\textwidth}{!}{%
\begin{tabular}{|c|c|}
\hline
Encoder                                                    & \begin{tabular}[c]{@{}c@{}}Implicit and non-linear \\ $ \hat{c} = \argmin\limits_{c} \sum_{(n,y)\in\omega} \Big(y- D_{\delta }\big( \thec,n\big) \Big)^{\thep}$ \end{tabular} \\ \hline
Decoder                                                    & \begin{tabular}[c]{@{}c@{}}Analytic and non-linear \\ $ \hat{y} = D_{\delta }\big( \thec,n\big) $\end{tabular}   \\ \hline
\begin{tabular}[c]{@{}c@{}}Compression (training)\\ step\end{tabular} & \begin{tabular}[c]{@{}c@{}} Optimization w.r.t. $(\delta, c)$ \\ $\min_{\delta, C}
\sum_{\omega\in\Omega}\sum_{(n,y)\in\omega}
\Big(y-D_{\delta}\big( C_{\omega},n\big)\Big)^{\thep} $\end{tabular} \\ \hline
\begin{tabular}[c]{@{}c@{}}Reconstructed surface \\ Reconstruction\end{tabular} & \begin{tabular}[c]{@{}c@{}}Implicit \\ $ \hat{y} = D\left( \argmin\limits_{c} \sum_{(n,y)\in\omega} \Big(y- D_{\delta }\big( \thec,n\big) \Big)^{\thep} , n \right)$ \end{tabular}  \\ \hline
\begin{tabular}[c]{@{}c@{}}Completed \\ surface\end{tabular} & \begin{tabular}[c]{@{}c@{}} $ \hat{y} = D\left( \argmin\limits_{c} \sum\limits_{{(n,y)\in\omega}} \Big(y- D_{\delta }\big( \thec,n\big) \Big)^{\thep} , n \right)$ \end{tabular}   \\ \hline
\end{tabular}%
}
\caption{The functional approach. }
\label{tab:Summary}
\end{table}


\begin{table}[!htbp]
\resizebox{\textwidth}{!}{%
\begin{tabular}{|c|c|c|}
\hline 
 & PCA  & Convolutional \tabularnewline
\hline 
Encoder  & %
\begin{tabular}{@{}c@{}}
Analytic and Linear\tabularnewline
$\hat{c}=E_{\varepsilon}(y)$\tabularnewline
\end{tabular} & %
\begin{tabular}{@{}c@{}}
Analytic and non-linear\tabularnewline
$\hat{c}=E_{\varepsilon}(y)$\tabularnewline
\end{tabular}\tabularnewline
\hline 
Decoder  & %
\begin{tabular}{@{}c@{}}
Analytic and linear\tabularnewline
$\hat{y}=D_{\delta}\big(\thec\big)$\tabularnewline
\end{tabular} & %
\begin{tabular}{@{}c@{}}
Analytic and non-linear\tabularnewline
$\hat{y}=D_{\delta}\big(\thec\big)$\tabularnewline
\end{tabular}\tabularnewline
\hline 
\begin{tabular}{@{}c@{}}
Compression (training)\tabularnewline
step\tabularnewline
\end{tabular} & %
\begin{tabular}{@{}c@{}}
Optimization w.r.t. $(\delta,\varepsilon)$\tabularnewline
$\min_{\theta=(\delta,\varepsilon)}\sum_{\omega\in\Omega}\sum_{(n,y)\in\omega}\left(y-\Big(D_{\delta}\big(E_{\varepsilon}(\omega)\big)\Big)_{n}\right)^{\thep}$\tabularnewline
\end{tabular} & %
\begin{tabular}{@{}c@{}}
Optimization w.r.t. $(\delta,\varepsilon)$\tabularnewline
$\min_{\theta=(\delta,\varepsilon)}\sum_{\omega\in\Omega}\sum_{(n,y)\in\omega}\left(y-\Big(D_{\delta}\big(E_{\varepsilon}(\omega)\big)\Big)_{n}\right)^{\thep}$\tabularnewline
\end{tabular}\tabularnewline
\hline 
\begin{tabular}{@{}c@{}}
Reconstructed surface \tabularnewline
Reconstruction\tabularnewline
\end{tabular} & %
\begin{tabular}{@{}c@{}}
Explicit/analytic\tabularnewline
$\hat{y}=D\left(E(y)\right)$\tabularnewline
\end{tabular} & %
\begin{tabular}{@{}c@{}}
Explicit/analytic\tabularnewline
$\hat{y}=D\left(E(y)\right)$\tabularnewline
\end{tabular}\tabularnewline
\hline 
\begin{tabular}{@{}c@{}}
Completed \tabularnewline
surface\tabularnewline
\end{tabular} & %
\begin{tabular}{@{}c@{}}
$\hat{y}=D\left(\argmin\limits _{c}\sum\limits_{{(n,y)\in\omega}}\Big(y-D_{\delta}\big(\thec\big)\Big)^{\thep}\right)$\tabularnewline
\end{tabular} & %
\begin{tabular}{@{}c@{}}
$\hat{y}=D\left(\argmin\limits _{c}\sum\limits_{{(n,y)\in\omega}}\Big(y-D_{\delta}\big(\thec\big)\Big)^{\thep}\right)$\tabularnewline
\end{tabular}\tabularnewline
\hline 
\end{tabular}}
\caption{PCA and convolutional approaches. }
\label{tab:Summary2}
\end{table}

Also note that, from a numerical complexity point of view, the functional approach is less sensitive to the dimension than, say, a classical autoencoder on a fixed grid (including our convolutional approach), for which the size of the grid typically grows exponentially with the dimension.
 


\section{Experimental Methodology and  Setting}\label{s:exp}

The data and code for the equity and repo case studies can be found on a public github repository \url{https://github.com/mChataign/smileCompletion} (the data and code for the swaption case study are proprietary).

In this section,
we devise an experimental methodology and the learning procedures, so that all models are set on comparable grounds.

All the optimization (compression or completion) problems are solved with the Adam adaptive learning rate stochastic gradient algorithms of \citeN{kingma2014adam}.
The output of a neural network is by construction non-convex with respect to its parameters. So are therefore all our loss functions.   
The Adam algorithm has proven its robustness in non-convex optimization context.
With the help of automatic adjoint differentiation, it provides fast training for most  neural networks architectures. 
However, no convergence is guaranteed theoretically. 
 
For the compression stage, we make a $80:20$
split of a full data set into a training set and a test set. 
The split is chronological in order to avoid look-ahead bias (cf.~\citeN{ruf2019neural}).
The training set is further split into a calibration and a validation data set.
The former is used for computing the gradients driving the numerical optimization in the training problem, 
whereas the latter is used for determining an early stopping rule that provides implicit regularization, as detailed below.

The learning rate of the Adam optimizer is set to $0.001$. 
Mini-batch learning is used in the repo and equity index derivative case studies, whereas batch-learning is employed with swaption volatilities.
The gradient descent is driven by the loss computed on the calibration set, but the validation error is the loss function computed on the validation data set.
The learning procedure is stopped when we do not observe any decrease of the validation error during a certain number of iterations, called patience.
The parametrization returned by the compression is the one that minimizes the validation error.
Early stopping in this sense limits the generalization error  (cf.~\shortciteN{EnglHankeNeubauer96}), i.e.~the gap between the reconstruction errors computed on the calibration data set and a new, unobserved data set, the role of which is played by the 
test set. Sometimes, as detailed later, a penalization term is added to the compression loss function in order to provide a more regular and stable minimization. 
A maximum number of iterations is fixed to $10^4$  at compression stage and $10^3$  at completion stage, in order to cap the length of the optimizations.
 
All approaches are implemented in Python, using the tensorflow package in the swaption case study and pytorch in the two others. Note that all hyperparameters are chosen manually, rather than by grid search or random search techniques. 
Grid search is not possible because we have too many hyperparameters. Exploring different neural net architectures would be too demanding computationally. However,
some of the hyperparameters can be fixed based on human expertise. 
For instance, 15 factors in our case study of Section \ref{s:eq} is the number of factors that equity derivative traders commonly use in PCAs (after interpolation on a fixed grid, as they are faced with moving grids). 
 
\subsection{Performance Metrics}


We want to assess, for each approach, the performance of the corresponding compression and completion procedures, as well as the behavior (distribution and dynamics) of the resulting factors. 
For the compression, we consider the average root mean square reconstruction error $RMSE_{\omega}$ on the test set $\Omega'$
(root mean square error $RMSE_{\omega}$
between the values at the nodes of the tensor $\omega$
and their reconstructed counterparts, i.e. 
\begin{equation}\label{eq:etdun} 
\sqrt{\frac{1}{\point} \sum\limits_{(n,y) \in \omega} \left(y -  \Big(D_{\delta^\star} \big(E_{\varepsilon^\star}(\omega)\big)\Big)_n \right)^2}\end{equation}
(or the analogous quantities with $m_\omega$ and $ D_{\delta^\star}\big( C^\star_{\omega},n\big)$ instead of $m$ and $\Big(D_{\delta^\star} \big(E_{\varepsilon^\star}(\omega)\big)\Big)_n$, as relevant).
In the case of the functional approach the encoder $E$ is implicit and its definition is detailed in table \ref{tab:Summary}.
We refer to \eqref{eq:etdun} 
as the reconstruction loss in the compression stage of our case studies given that $\omega$ is a complete surface.

In constrast with \eqref{eq:etdun},
\begin{equation}\label{eq:completionLoss}
\sqrt{\frac{1}{\point} \sum\limits_{(n,y) \in \omega} \left(y - \Big(D_{\delta^\star} \big(\hat{c}\big)\Big)_n \right)^2}\end{equation}
(or $m_\omega$ rather than $m$ in the case of the functional approach)
is called the completion loss when we compare the complete original observation with the completed observation. This completed observation is given by the decoder for code values $\hat{c}$ which are calibrated on the incomplete view provided by $\omega$.

In the case of interpolation benchmarks, there is no compression stage and no code is involved at the completion stage: the completion loss
is then defined
by the RMSE between the interpolated surface (from an incomplete $\omega$) and the original complete $\omega$.
 
We provide a focus on the observation $\omega$ leading to the worst $RMSE_{\omega}$ over the test set, in order to identify the locations
that are less well handled (e.g.~short option maturities).
In addition,
we display the time series of the codes.
A good compression should exploit each factor in the code (we should not observe factors stuck at zero).

The quality of the completion is assessed by a backtest on the test set. Each day of $\Omega'$, we solve the problem \eqref{eq:Completion} or \eqref{eqCompletion1}, initialiazing the factors with the fully informed encoding of the previous day.
We then mask 90 \% of the points in each tensor of the test set. 
For each such observation $\omega\in\Omega'$, we check the reconstruction $RMSE_\omega$ between the completed surface and the true one.
Like for compression, we plot the worst completion obtained on the test set $\Omega'$.

\subsection{Introduction to the Case Studies}\label{ss:casestud}

We provide numerical results on three daily time series of real financial data:
repurchase agreement yield rates, equity implied volatility surfaces and at-the-money swaption implied volatilities.
However, the swaption implied volatilities have been preprocessed by our data provider
to fit a fixed grid (whereas the native, raw data had a moving time-to-expiry). A preprocessing entails
an unquantifiable
bias and our recommendation would be to  apply the functional approach to the original data (whenever available).
The main motivation for the third example
is that one can then benchmark the functional approach against PCA and the convolutional approach. 

The advantage of working with yield rates or implied volatilities, instead of the corresponding option prices, is that these are scaled quantities, exempt from first order dependence on contract characteristics such as nominal, time-to-maturity, actual level of the underlying  in at-the-money option data, etc., which should otherwise be added to the set of explanatory variables in all learning procedures. 
The ensuing arbitrage issue is discussed in the next subsection.
 

\subsection{Discussion of the Arbitrage Issue}
Arbitrage constraints can be expressed naturally in terms of options prices using calendar spread and butterfly. But in terms of implied volatility, they are non-trivial, even in the simplest case of equity derivatives (for which they are fully stated in 
\citeN{roper10}). No compression/completion method applied to implied volatility surfaces provides a way to deal with those constraints without coming back inherently to option prices. In order to circumvent that problem, one could apply our approach to the coefficients of a (e.g. local vol) model, from which non arbitrable prices and implied volatilities could be derived in a second step.
 However, we do not choose this route because: 
\begin{itemize}
\item the market practitioners, who play both the roles of human experts and users, have built intuitions over decades on implied volatilities. They think of option prices directly in terms of implied volatilities. Providing them with a good recommendation tool in terms of a quantity that is familiar to them is of great value and the primary purpose of our approach; 
\item most of the times, the starting point for calibrating a model (e.g. Dupire) is nothing else than the implied volatilities. Therefore the trader must correct the anomalies {\em before} the implied volatility surface can be plugged as an input to model calibration. Hence one of the requirements of our proposed approach is that it should be model-free;
\item  Having said this, if one assumes that, on the one hand, most of the surfaces in our database are arbitrage-free and, on the other hand, a more regular surface is less prone to arbitrage opportunities, then one concludes that our model should  tend to remove part of the arbitrages present in the data. This can actually be seen empirically on some of the examples in Section \ref{s:eq}. This is a natural by-product of anomaly correction and it also eases the calibration process.
\end{itemize}
Similar comments apply on most markets (beyond equity implied volatility), including the ones of our three case studies,  i.e. repo
contracts, handled 
by traders
 in terms of 
  yield curves, and equity index derivatives and swaptions, which are handled in terms of implied volatility.

\section{Repo Curves}\label{s:repo}

Our first case study bears on the nowcasting of repo rates, based on an 2013--2019 daily time series of repo yield curves (repo rates, where repo is a shorthand for repurchase agreement).

The grid of nodes in the data is  unstructured, in the sense that the corresponding {dates (time-to-maturities of bonds with standardized maturity dates)} 
vary, in both number and location, from day to day  (with as little as two or three points on particularly idle days),
see e.g. Figure \ref{fig:lier}. 
Indeed, as the expiration dates used to compute the repo curve are fixed, and the variable of interest for the repo curve shape is rather time to expiry, the latter decreases as the expiry date approaches. 
For a given repo curve, the times to expiry for which the repo value is available is not known in advance for that reason. 
Therefore, there is no canonical way to have a systematic representation of repo curves on a fixed grid, one would need to introduce artificial time to expiry of interest and interpolate/extrapolate (which poses issues of its own) the repo curve to get the values, and then working on transformed data.
This is the situation the functional approach is tailored for.
 By not making any assumptions on the domain of input (time to expiry), the functional approach enables to handle unaltered data, by
 treating the {time-to-maturity} of a transaction as an input value 
(cf.~Figure  \ref{fig:jiangNN}).

\subsection{Functional Network Architecture}\label{ss:ftuning}
\begin{figure}[!htbp]
  \centering
  \includegraphics[width=\linewidth]{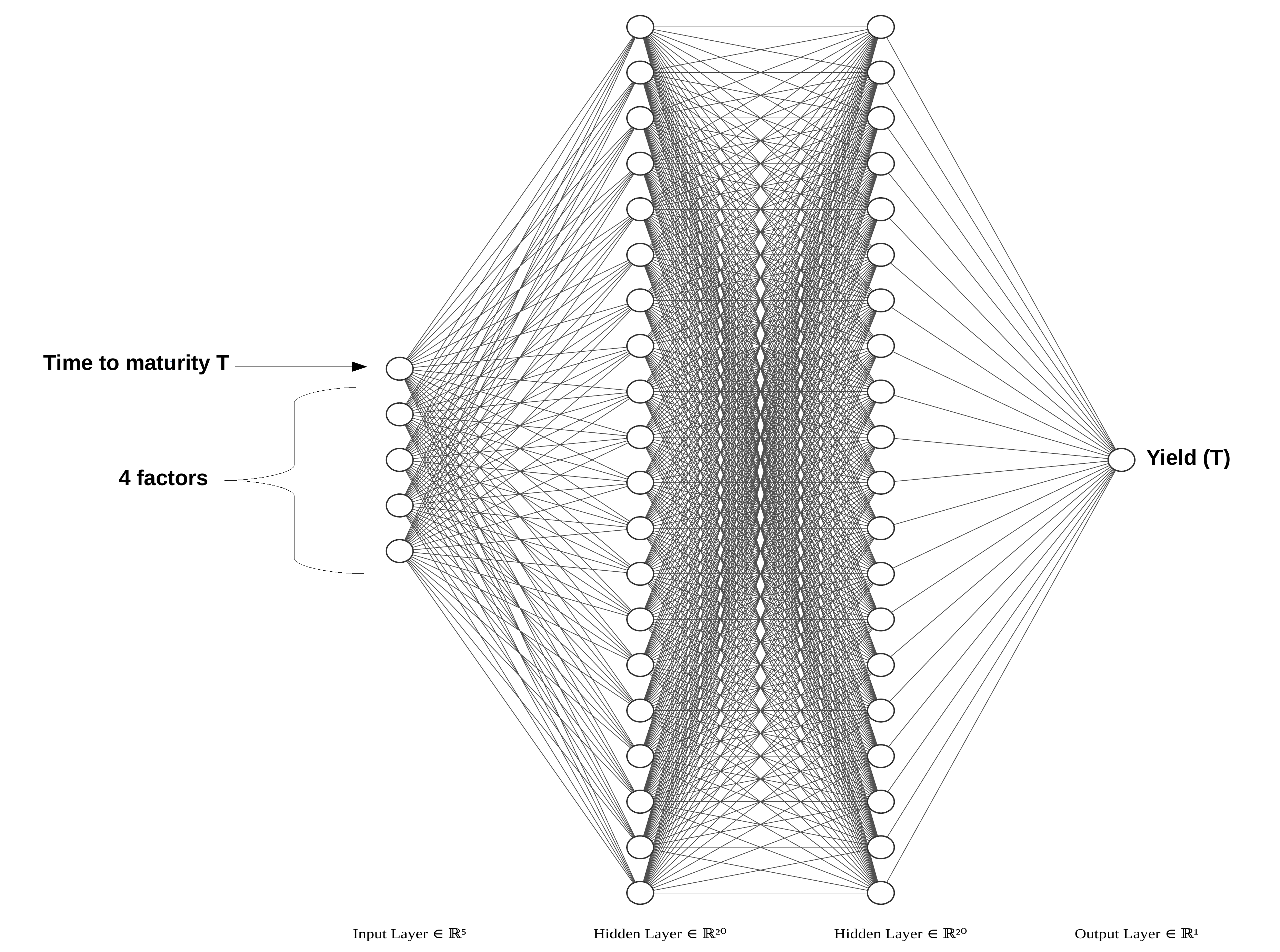}
  \caption{Network of the \functional approach used in 
  the repo
  case study. Here and in Figures  \ref{fig:jiangNNeq} and \ref{fig:jiang} below, the
  graphs have been produced using the style FCNN of the NN-SVG software: the units and the connections between them are represented by circles and edges.} 
  \label{fig:jiangNN} 
\end{figure}

Our \functional approach is implemented by a single feed-forward neural network 
composed of three 
fully-connected layers with 20, 20 and 1 units  (see Figure \ref{fig:jiangNN}).
Hyperbolic tangent activation is applied to each but the output layer for the same reasons as above (and the output layer is linear).

 \subsection{Numerical Results}
 
\begin{figure}[htbp]
  \centering
  \includegraphics[width=0.45\linewidth]{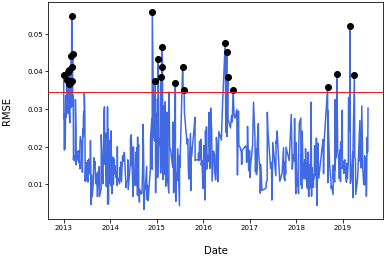}  \includegraphics[width=0.45\linewidth]{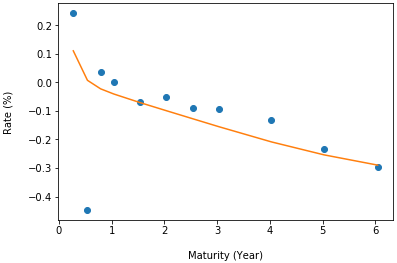}\\
  \includegraphics[width=0.45\linewidth]{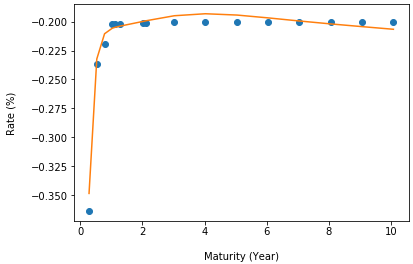}  \includegraphics[width=0.45\linewidth]{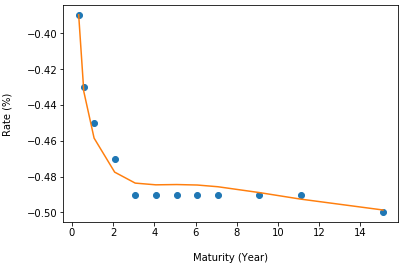}

  \caption{{\em (Bottom)} Interpolation of two inlier repo curves; {\em (Top left)} Time series of the (absolute) RMSEs on the repo data and 0.035 RMSE threshold; the spotted values correspond to the outliers at the chosen threshold. {\em (Top right)} Interpolation of an outlier repo curve.}
  \label{fig:lier} 
\end{figure}

 As the bottom panels of Figure \ref{fig:lier} illustrate,
the parameterization is flexible and can accomodate different curve shapes or node localizations.

As explained in Section \ref{ss:outl-detn}, the compression stage can be used for detecting an abnormal curve and correcting it with a more likely one.
The distinction between inliers and outliers is determined by a threshold on the reconstruction error.
A bad reconstruction is taken as a signal that the codebook is not able to explain the corresponding observation. We then conclude that the latter does not lie {in the manifold $\mathcal{S}$ of the ``usual'' curves, hence we classify it as an outlier  (see Section \ref{ss:outl-detn}).} We can then correct (replace) these data by the curve reconstructed from the decoder with the factors calibrated on the current values, i.e. by the output of the corresponding completion
\eqref{eqCompletion1}. 


 The lower panels of Figure \ref{fig:lier} show the gap between the observed data points and the reconstructed ones. The upper left panel spots the outliers at 
 a 0.035 absolute RMSE threshold. The upper right panel gives an example of outlier correction.
 
\section{Equity Derivative Implied Volatility Surfaces}\label{s:eq} 

As a second experiment, we apply our functional approach to Black--Scholes implied volatilities surfaces of equity index derivatives.
The corresponding volatilities price options on the Nikkei 225 index from 2015 to 2018 (included), corresponding to 1544 observable surfaces. The order of magnitude of implied volatilities fluctuates between 0.15 and 1.2.
We include the forward rate as an exogenous variable   that can be plugged into the functional network \eqref{fig:jiangNN} along with log-maturity and log-moneyness.

\begin{figure}[!htbp] 
  \centering
  \includegraphics[width=\linewidth]{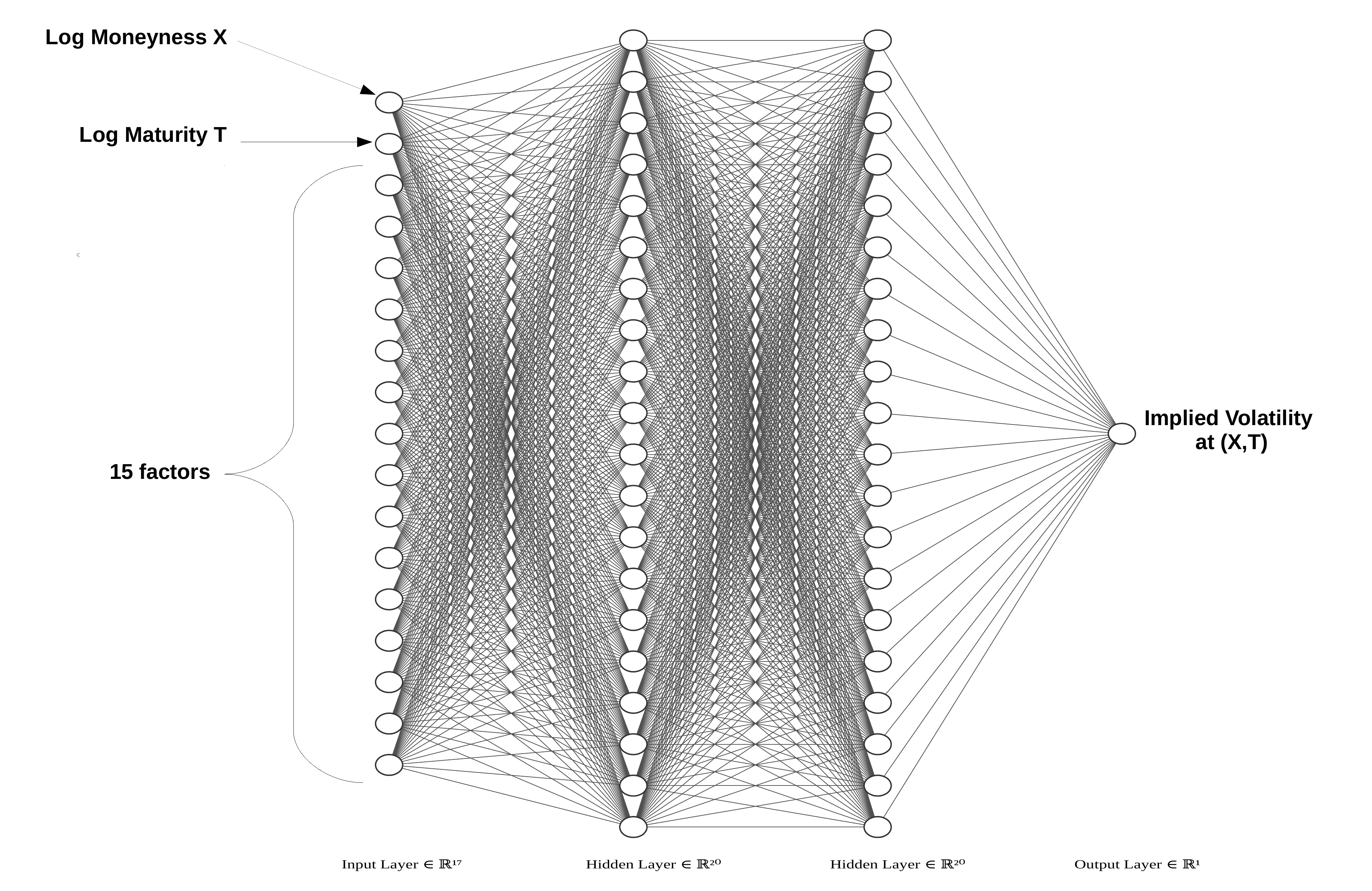}
  \caption{Network of the \functional approach used in
  the equity
  case study (style FCNN of the NN-SVG software, cf. Figure \ref{fig:jiangNN}).}
  \label{fig:jiangNNeq} 
\end{figure}

As in the repo case study,
the grid of nodes in the data is unstructured, in the sense that the corresponding dates (time-to-maturities of equity index options with standardized maturity dates)
But, again, this is the situation the functional approach is tailored for  (cf.~Figure  \ref{fig:jiangNN}). 
The corresponding architecture of the \functional approach is then similar to the one used for repo curves in the previous section,
except that the log-time-to-maturity and the log-moneyness are used as the (two dimensional) localization inputs, and that 15 latent variables are used (instead of only 4 previously):  see Figure \ref{fig:jiangNNeq}.  
Moreover, 
one can also easily incorporate the forwards 
 as exogenous variables.
For taking them into account, it suffices to add to the network
of Figure \ref{fig:jiangNNeq} an additional feature (input unit) containing the level of the  forward swap rate with maturity $T$.
Hence,
the units for the
maturity $T$
indicate the common location of the corresponding 
volatilities and forward 
rates.

\subsection{Compression}

We first calibrate our functional approach with the compression stage.
Toward this end, we execute the optimization \eqref{eq:Compression} on the training set and then calibrate codes with \eqref{eq:Completion} for each observation in both testing and training data sets. The quality of the
compression is assessed through the reconstruction errors reported in Table \ref{tab:compression}. By
reconstruction error we mean the gap between the original surface and the surface induced by the code  calibrated from \eqref{eq:Completion}.

We emphasize the difference between a reconstucted surface (as above)  and  a completed surface (considered later): the code leading to the completed surface is calibrated from an incomplete surface whereas the one for the reconstructed surface is obtained from a complete real surface.


\begin{table}[!ht]
\centering
\begin{tabular}{|c|c|c|}
\hline
 & Functional & \begin{tabular}[c]{@{}c@{}}Functional \\ with Forward\end{tabular} \\ \hline
Training set & 0.0070 & 0.0063 \\ \hline
Testing set & 0.0058 & 0.0064	 \\ \hline
\end{tabular}
\caption{RMSEs for reconstructed implied volatilities. }
\label{tab:compression}
\end{table}

In all four cases, the RMSEs in Table \ref{tab:compression} are very small compared to the order of magnitude of implied volatilities (between 0.15 and 1.2). The results show no sign of overfitting (the reconstructions error are similar on the training set and the testing set). Moreover the comparison between the two columns of the table
indicates that there is no benefit in including the forward price as an exogenous variable in our network.

\begin{figure}[!htbp]
\centering
\includegraphics[width=\linewidth]{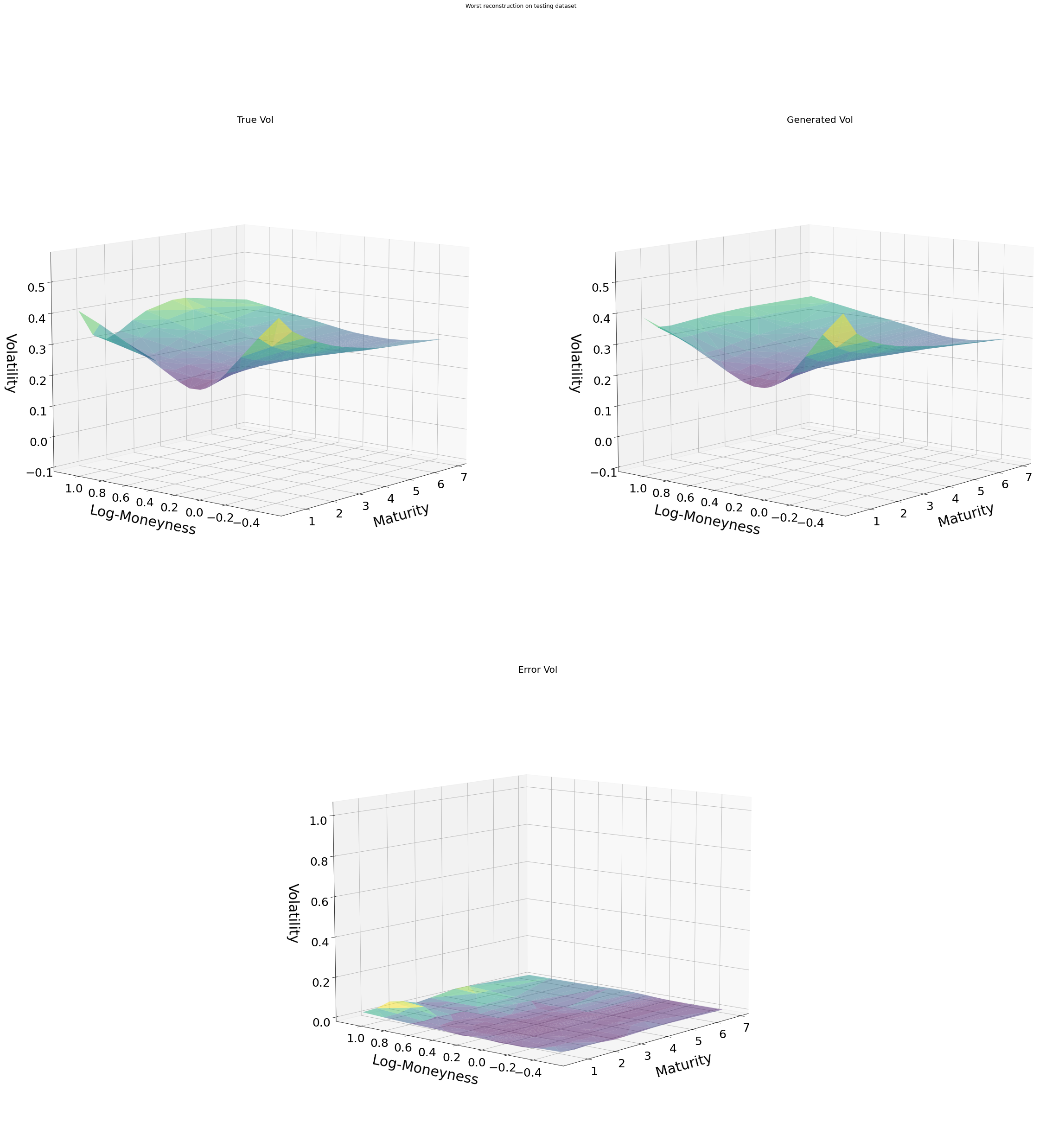}
\caption{Original surface vs compressed surface yielding worst RMSE.} 
\label{fig:compressionOutlier}
\end{figure}

Another way to assess the performance of the compression stage is to consider the worst compression, i.e. the surface yielding the highest reconstruction error.
This worst reconstruction corresponds to a RMSE of 0.0096. It is represented in Figure \ref{fig:compressionOutlier}, with the real surface on the top-left corner, the reconstructed couterpart on the top-right corner and the pointwise absolute difference between the two at the bottom.

We notice that the errors are concentrated on the upper tail (deep in the money call options) and for short maturities, which corresponds to illiquid options.

\begin{figure}[!htbp]
\centering
\includegraphics[width=.48\textwidth]{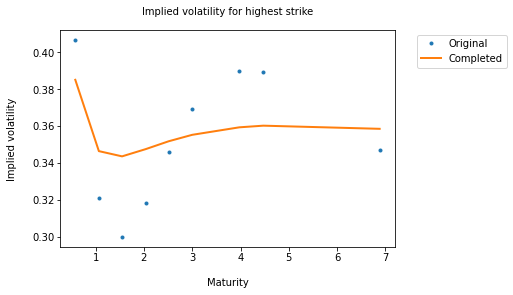}\quad
\includegraphics[width=.48\textwidth]{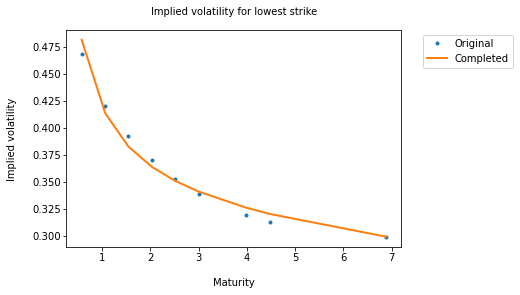}
\caption{Tails of compressed surface vs original implied volatilities.}
\label{fig:compressionOutlierTail}
\end{figure}

A bad reconstruction of a surface
can also be used for qualifying it as an outlier. For instance, 
Figure \ref{fig:compressionOutlierTail} shows the implied volatility values corresponding to the most extreme strikes in Figure \ref{fig:compressionOutlier}: original data points as dots and curves from the reconstructed surface.
The left panel corresponding to the illiquid upper tail shows around the maturity 1.5 year a very low point that an expert would indeed qualify as an anomaly.
The correction (i.e. the reconstructed surface) ignores this anomaly and has a more reasonable shape from a practitioner of view.

The left part of Figure \ref{fig:OutlierTheta} shows that the corrected surface is not prone to calendar arbitrage: the sensitivity to the maturity  of the corresponding implied total variance is positive for every maturity $T$.\footnote{Regarding butterfly arbitrages,
Durrleman's condition on the density (involving sensitivity with respect to forward log-moneyness, cf. \citeN{roper10}) can unfortunately not be checked for lack of data regarding dividends and discounting.} Sensitivity is computed thanks to adjoint automatic differentiation from neural  network.

The above example shows that the functional neural network is indeed apt to learn from the compression stage  a low-dimensional representation of likely observations.
The low-dimensional representation gives large reconstruction errors to
the surfaces of the testing set atypical with respect to the past observations (the training set in our experiments)
and their latent structure.

\subsection{Outlier Detection and Correction}\label{ss:odc}

\begin{figure}[!htbp]
\centering
\includegraphics[width=.48\textwidth]{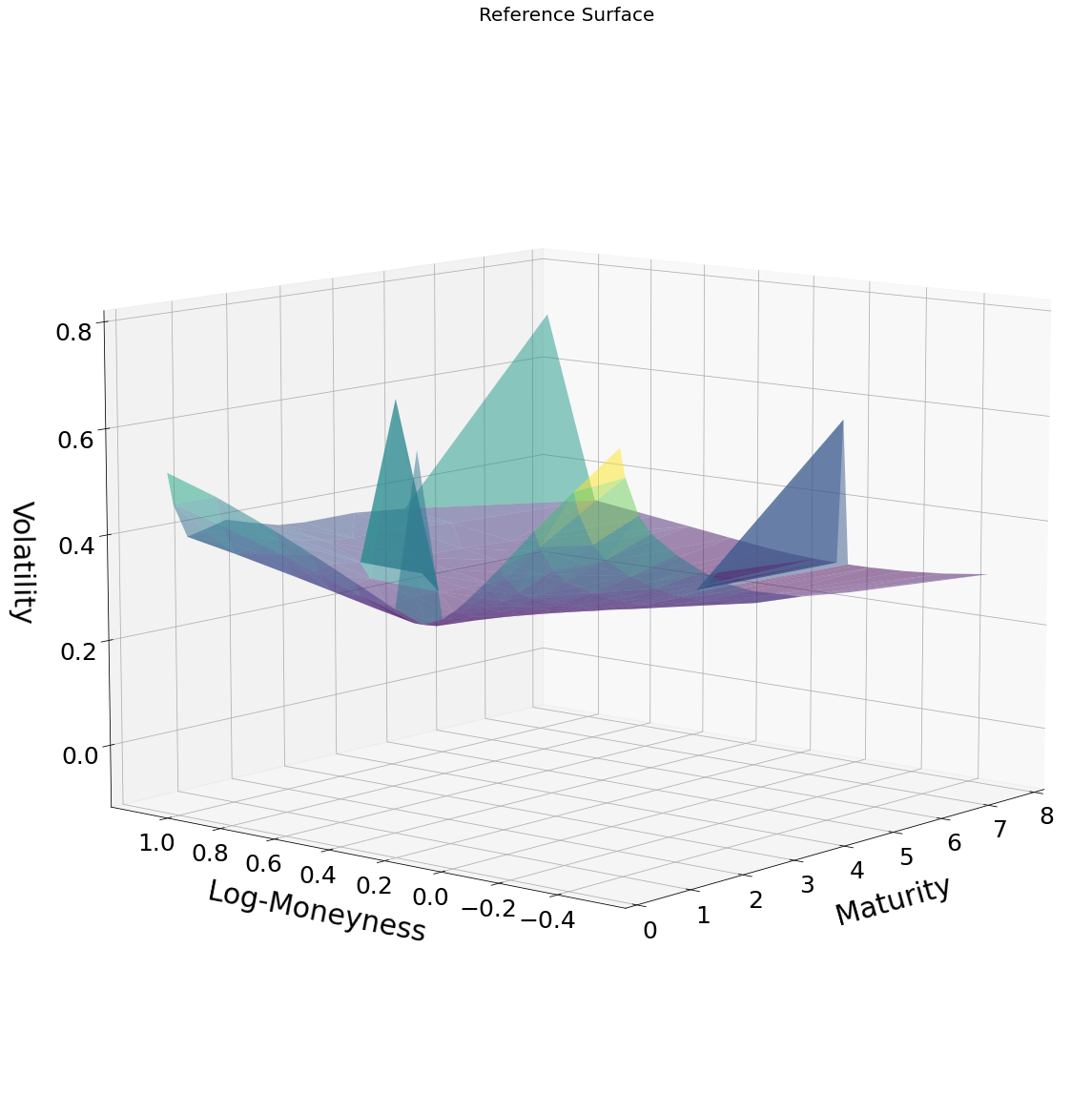}\quad
\includegraphics[width=.48\textwidth]{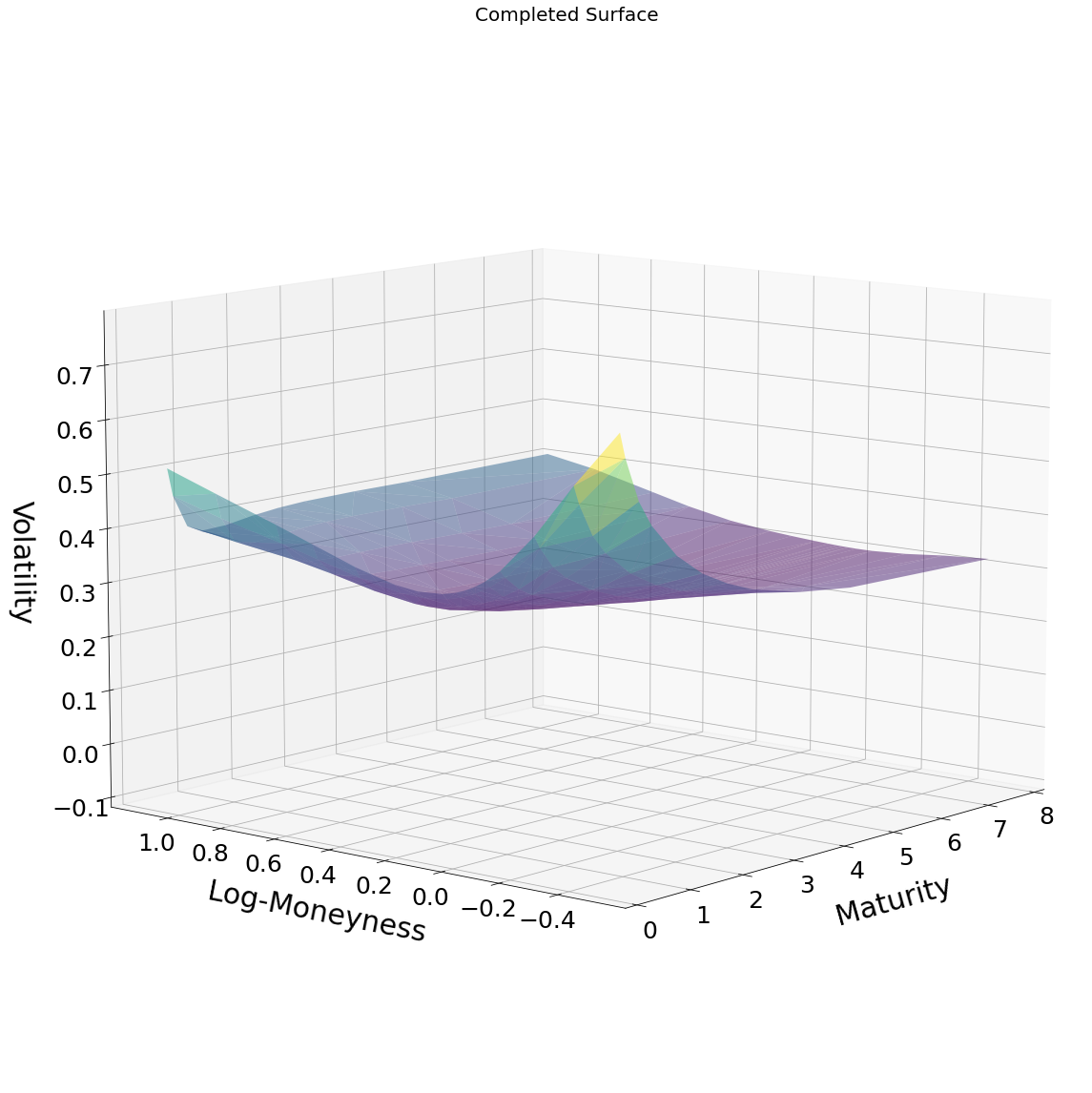}\\
\includegraphics[width=.48\textwidth]{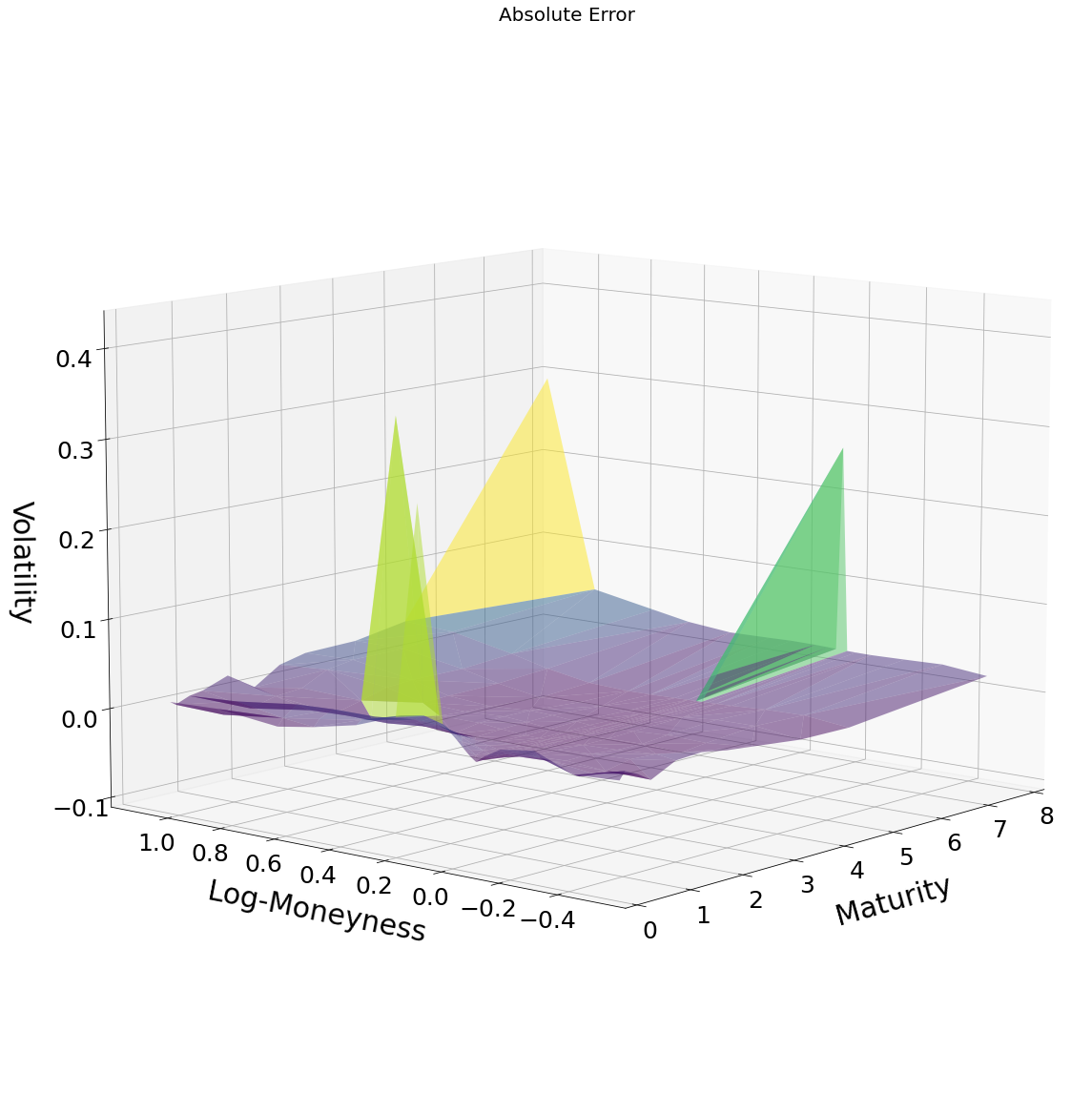}\quad
\includegraphics[width=.48\textwidth]{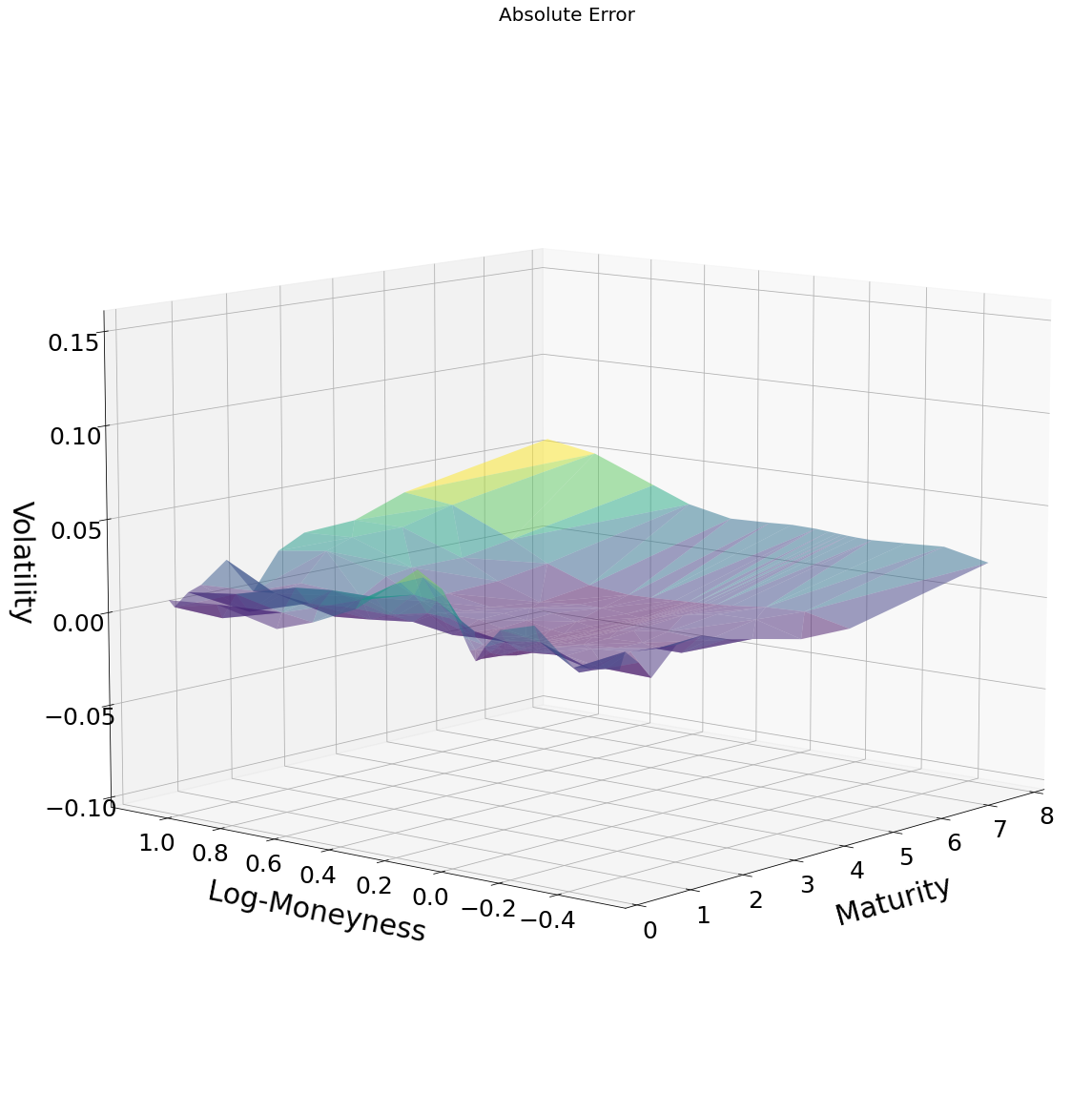}
\caption{Outlier correction : Corrupted surface (Top-left), Corrected surface (Top-right), absolute error between corruption and correction (bottom-left), absolute error between correction and original surface before corruption (bottom-right)}
\label{fig:dummyOutlier}
\end{figure}

To confirm our views on outliers, we propose the following sanity check.
An observation (first volatility surface in the testing set) is chosen and articially corrupted by doubling the values on four randomly chosen points:
see the top-left corner in Figure \ref{fig:dummyOutlier}.

Then we run the optimization \eqref{eq:Completion} on this corrupted surface and obtain recalibrated codes.
These code produce with the decoder the reconstructed surface (called correction) on the top-right corner.
The correction is a smooth surface in which the corrupted values have been overwritten  by values close to the original (non corrupted) ones.
The bottom-left panel shows that only the corrupted values have been modifed significantly by the correction stage. The bottom-right figure indicates that the corrected surface is very close to the original one. 
The RMSE between the  corrupted and the corrected surface is $0.0446$ whereas the one between the correction and the original surface is $0.0151$.

\begin{figure}[!htbp]
\centering
\includegraphics[width=.48\linewidth]{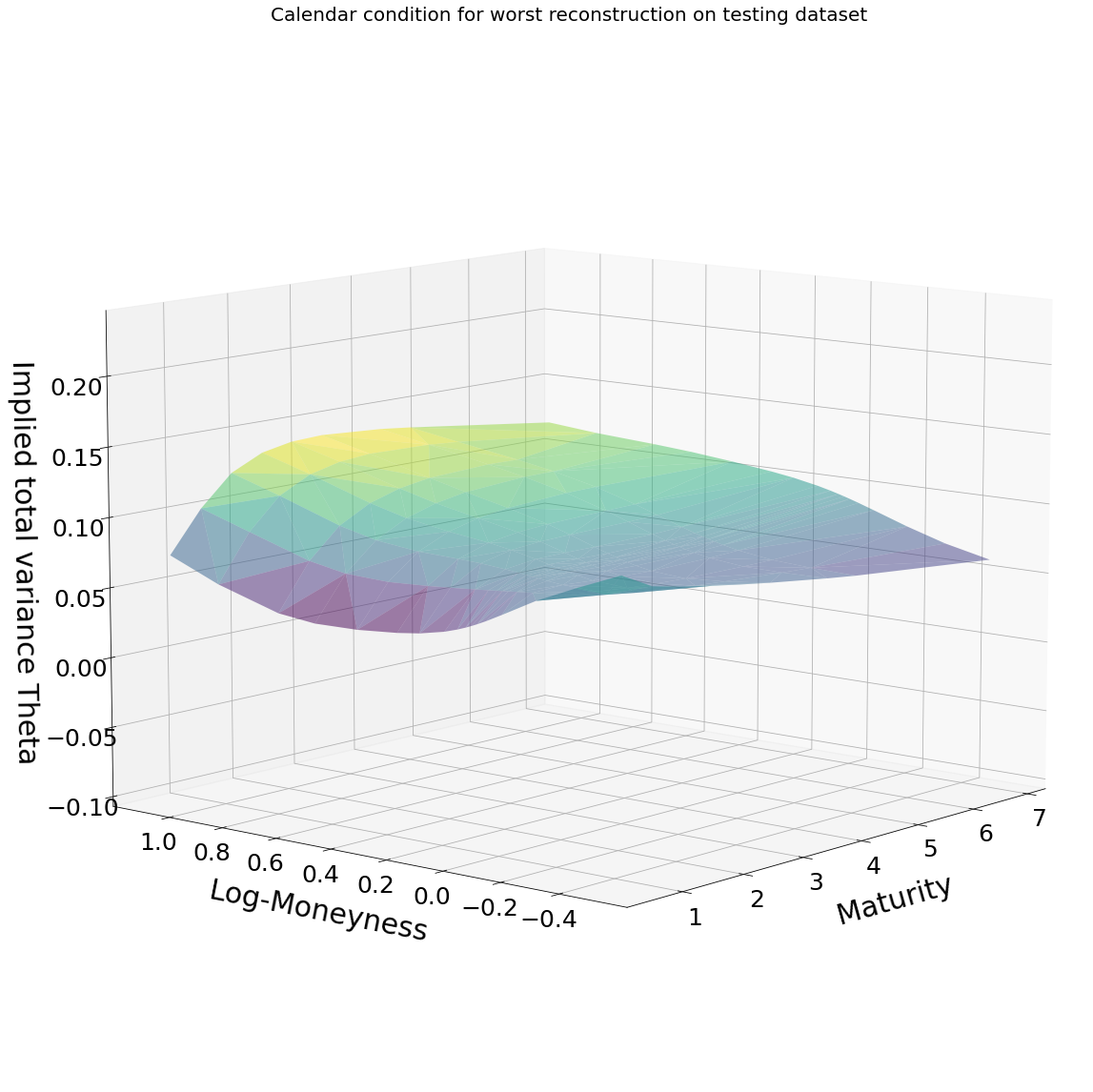}\quad
\includegraphics[width=.48\linewidth]{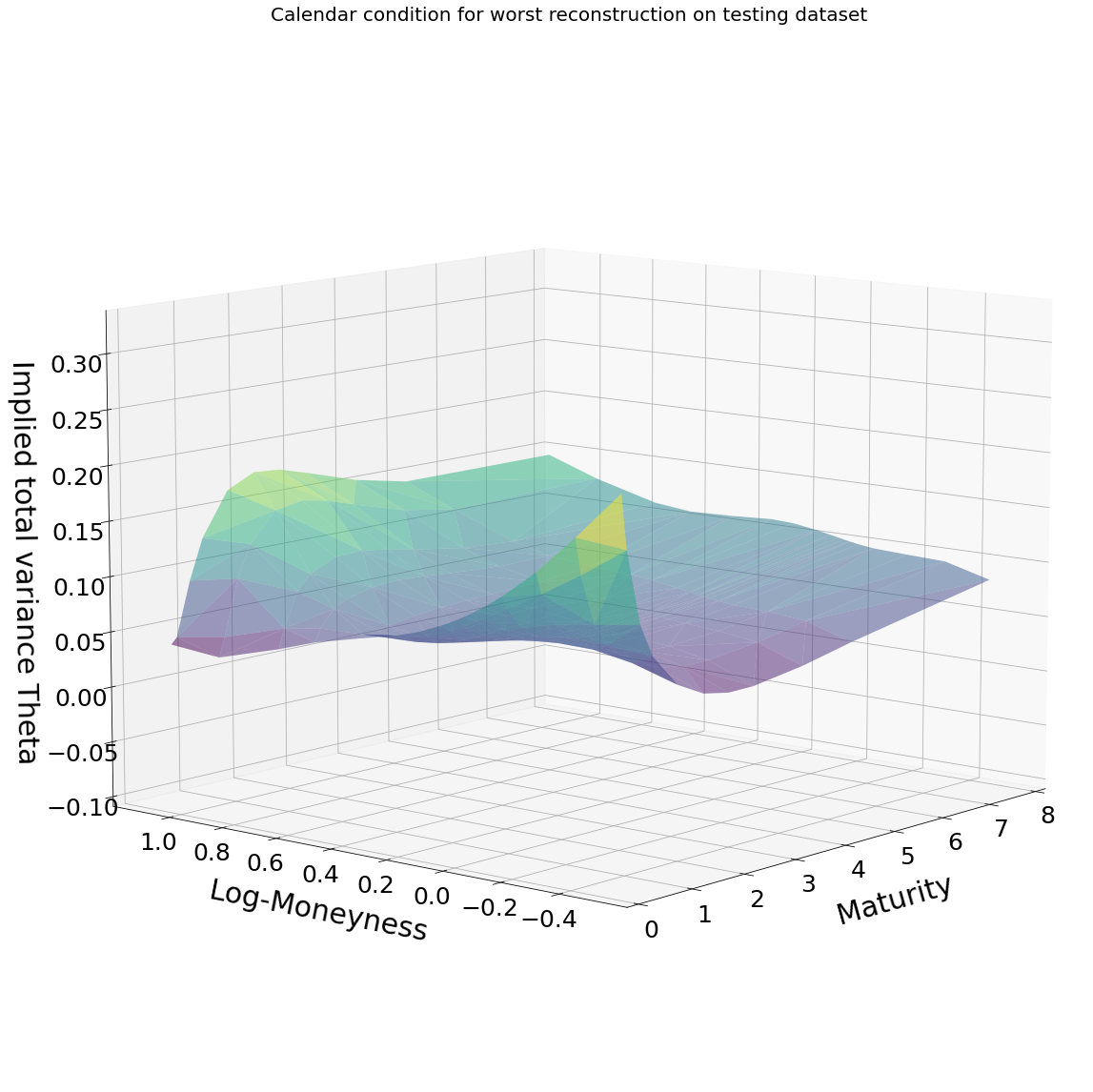}\\
\includegraphics[width=.48\linewidth]{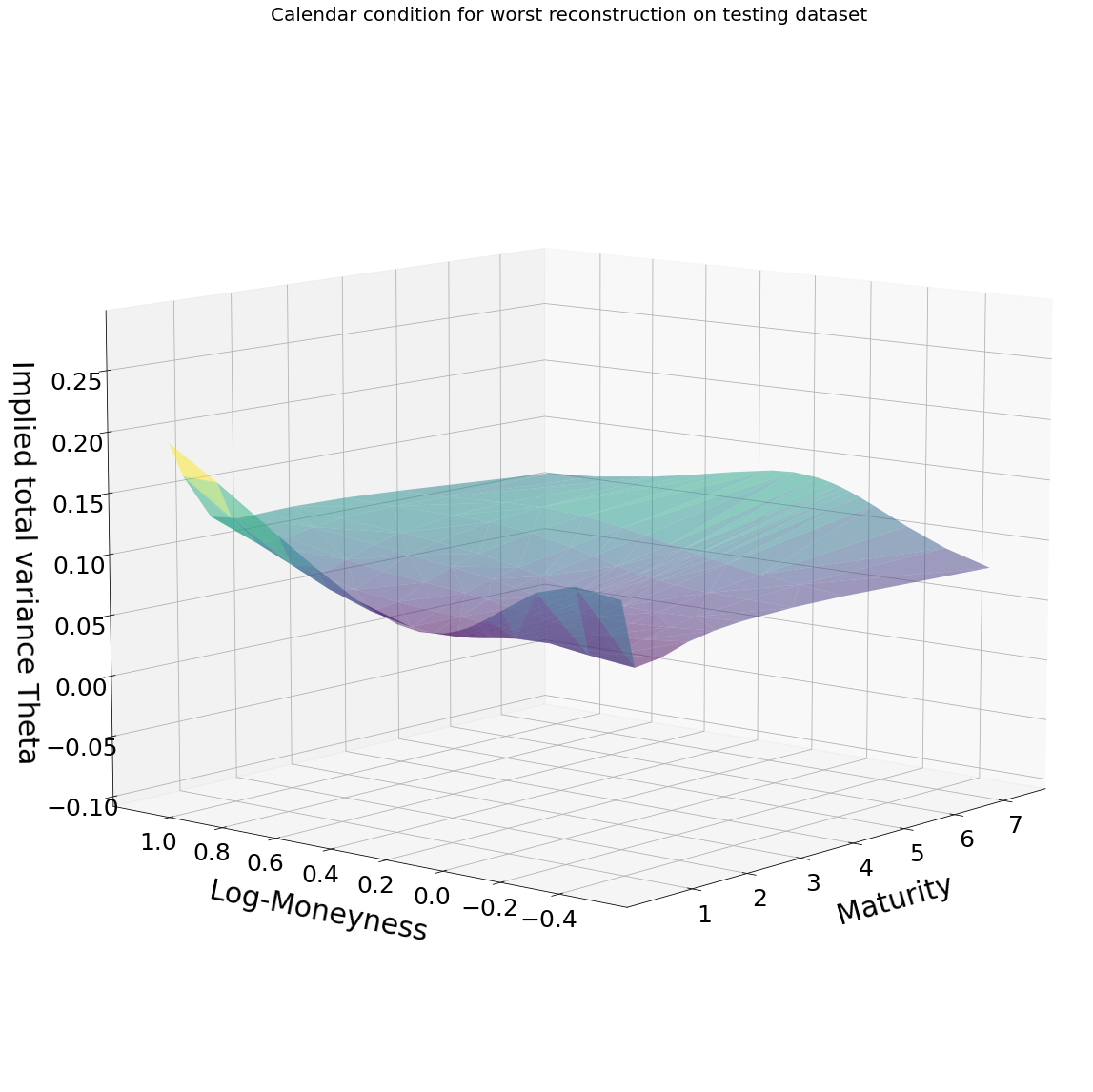}
\caption{Implied total variance theta for worst reconstruction on top-left, outlier correction on top-right and worst completion at the bottom.}
\label{fig:OutlierTheta}
\end{figure}

Note that the calendar arbitrage condition is still respected (see figure \ref{fig:OutlierTheta}) for the correction, which exhibits a positive sensitivity of the implied total variance with respect to the maturity of the option. 

This experiment confirms that a high reconstruction error is a good indicator of an outlier. The calibrated latent structure of the functional network smoothes 
the corresponding 
surface
 by identifying and correcting its anomalous   points. 

\subsection{Completion}

We now want to leverage on the calibrated low-dimensional latent structure of the functional network to recover a complete surface from partial information.
Our hope is that this procedure will generate likely surfaces while approaching the available values (including on moving grids).

For each observation (surface) in the testing set, we select 40 points among the 255 points and remove all the others.
Then we calibrate the latent variables by solving numerically the problem \eqref{eq:Completion} with loss corresponding to these 40 points.

In order to benchmark the functional approach and assess the contribution of the historical data to the performance of the method,
 we report average completion errors \footnote{Gap between the original surface and the  completed one.} on the testing set for standard interpolation procedures (within each given surface, without exploitation of the information provided by the others):
\begin{enumerate}
    \item Linear interpolation: given a triangulation of the 2D maturity and log-moneyness space base on the locations of the 40 available points, the interpolated value is taken as the barycenter on each triangle;
    \item Spline interpolation: uses in each triangle as above a piecewise cubic interpolating Bezier polynomial (see \citeN{alfeld1984trivariate} and the scipy documentation of the CloughTocher2DInterpolator method);
    \item Gaussian process regression and squared exponential kernel: denoting by $X$ the observed locations (maturity and log-moneyness), by $Y$ the observed lognormal volatilities at locations $X$, by $X^{\star}$ the locations without values and by $Y^{\star}$ the unknown (looked for) implied volatilities, a Gaussian process regression assumes a Gaussian distribution  \begin{equation}\small\label{e:GPR}(Y, Y^{\star}) \sim \mathcal{N}(0, \begin{pmatrix} K(X, X) & K(X, X^{\star}) \\ K(X^{\star}, X) & K(X^{\star}, X^{\star}) \end{pmatrix} )\mbox{ with }   K(X, X^{\star})_{ij} = \sigma \exp{\left(- \frac{\|x_i - x_j\|^2}{l^2} \right)},\end{equation} where $\sigma$ and $l$ are two hyperparameters calibrated by log-likelihood to the  available values. 
    In \eqref{e:GPR}, $$\|x_i - x_j\|^2=\left(T_i - T_j\right)^2 + \left(\ln{(m_i)} - \ln{(m_j)}\right)^2 ,$$ where $T$ denotes a maturity and $\ln{(m)}$ a log-moneyness;
    \item Gaussian process regression with flat extrapolation; similar to 3, except that the  Gaussian process predictor is only used for interpolation purposes; extrapolation whenever required is performed by the nearest neighbour method.
\end{enumerate}

 Again, a major difference between our functional (or neural net more generally) approach
 and these interpolation benchmarks is that, in order to interpolate a given surface, the neural network takes into account the  information contained in all the surfaces of the data set, which is used as training set at the compression stage. 
In contrast, the above interpolation benchmarks only use the information provided by the 
available points of the currently interpolated surface, without consideration of the other surfaces in the data set. In particular, by Gaussian process regression in 3. and 4., we just mean interpolation within a given surface, using the available points in this surface as training set (unrelated to the potential use of Gaussian processes as an alternative to neural networks in our compression/completion approaches, which would be unrealistic as discussed in Subsection \ref{ss:compr}).

Accordingly, the functional approach exhibits significantly smaller completion errors. In Table \ref{tab:completion}, 
we reported these errors for two different choices of the 40 visible points : 
\begin{itemize}
    \item Less correlated points, i.e. locations for which the implied volatilities are the less correlated;
    \item Uniformly spread points, i.e.a random selection of at least 2 points per maturity. The lowest maturity can be assigned 3 visible points in order to reach a total number of 40 points. 
\end{itemize}
As the loss in \eqref{eq:Completion} is now computed on much fewer points (partial information in this sense),
the compression errors of the functional approach are obviously higher than the reconstruction errors from Table \ref{tab:compression}.
Smaller error are reported in the second case above because less correlated points are rather located on short maturities, so that, in the first case little information, is available  for the long maturities.

\begin{table}[!ht]
\centering
\resizebox{\textwidth}{!}{%
\begin{tabular}{|c|c|c|c|c|c|c|}
\hline
 & Functional & \begin{tabular}[c]{@{}c@{}}Functional \\ with Forward\end{tabular} & \begin{tabular}[c]{@{}c@{}}Linear \\ interpolation\end{tabular} & \begin{tabular}[c]{@{}c@{}}Spline \\ interpolation\end{tabular} & \begin{tabular}[c]{@{}c@{}}Gaussian process \\ no extrapolation\end{tabular} & \begin{tabular}[c]{@{}c@{}}Gaussian process \\ flat extrapolation\end{tabular} \\ \hline
Less correlated points & 0.0262 & 0.0265 & 0.0632 & 0.0462 & 0.0555	 & 0.0459 \\ \hline
Uniformly spread points & 0.0076 & 0.0091 & 0.0211 & 0.0168 & 0.0201 & 0.0208 \\ \hline
\end{tabular}%
}
\caption{RMSEs  for completed implied volatilities. }
\label{tab:completion}
\end{table}

All the completion results reported hereafter correspond to the case of uniformly spread visible points.

\begin{figure}[!htbp]
\centering
\includegraphics[width=.48\textwidth]{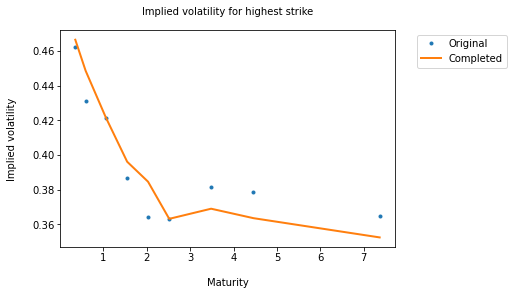}\quad
\includegraphics[width=.48\textwidth]{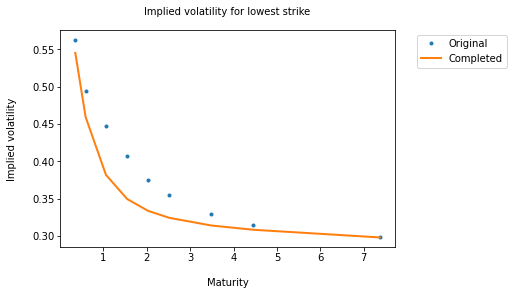}
\caption{Tails of completed surface vs. original implied volatilities.}
\label{fig:completionOutlierTail}
\end{figure}

The completion
method provided by the functional approach
is also robust: even the worst completion does not produce an outlier, i.e.  
\begin{itemize}
    \item the completed surface is smooth,
    \item the completed surface has a shape similar to the one of the original surface (the pointwise errors between the original and the completed surfaces are uniformly distributed), 
    \item the implied total variance sensitivity with respect to the maturity is still positive (see Figure \ref{fig:OutlierTheta}), inducing no calendar arbitrage opportunity,
    \item tails are consistent with the original points (see Figure \ref{fig:completionOutlierTail}) and not irregular.
\end{itemize}

\begin{figure}[!htbp]
\centering
\includegraphics[width=\linewidth]{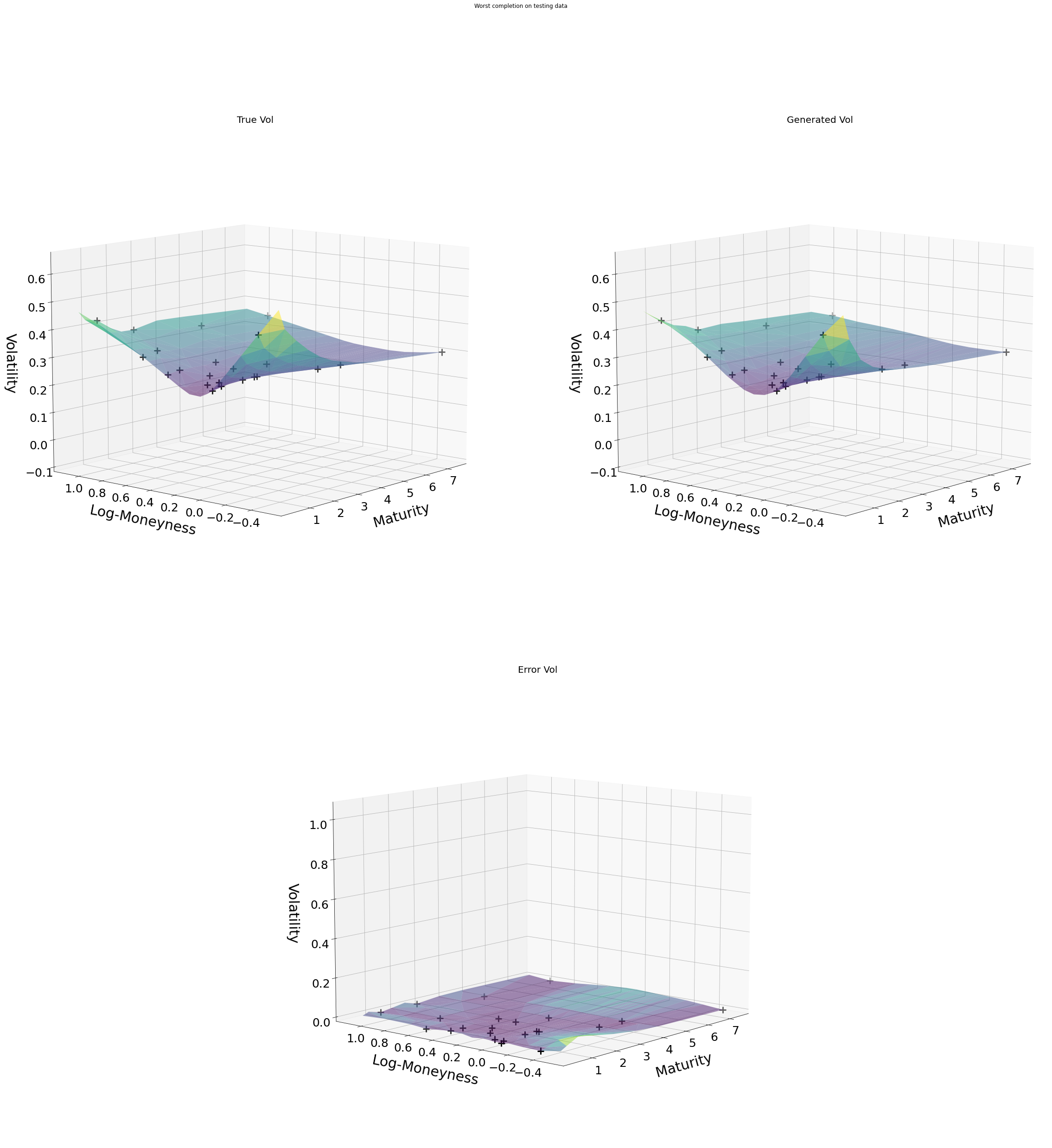}
\caption{Original surface vs. completed surface yielding the worst RMSE. Black crosses mark visible points.}
\label{fig:completionOutlier}
\end{figure}

\begin{figure}[!htbp]
\centering
\includegraphics[width=\linewidth]{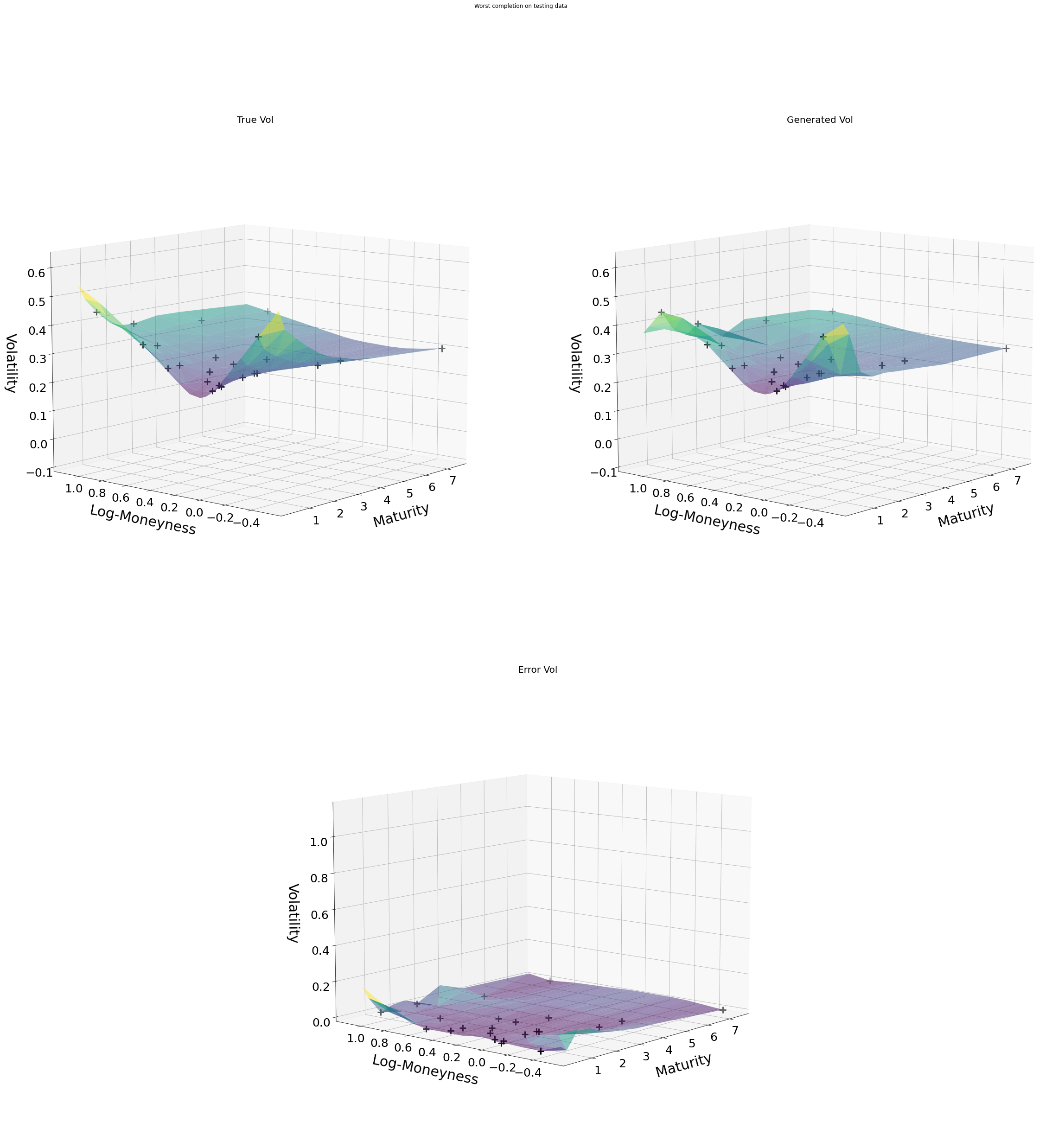}
\caption{Original surface vs. completed surface yielding the worst RMSE with spline interpolation. Black crosses mark visible points.}
\label{fig:completionOutlierLinear}
\end{figure}

Such robustness is not provided by the interpolation benchmarks.
For instance, in the case of the worst completion with the spline interpolation,
the completed surface (top-right corner of Figure \ref{fig:completionOutlierLinear}) is irregular in the tails.

\section{At-the-Money Swaption Surfaces}\label{s:swn}

The previous section was showing a case where the functional approach outperforms elementary interpolation benchmarks in an situation (in fact, the most  common in the context of financial nowcasting applications) involving a moving grid.

We now consider an application where the grid is constant (after a preprocessing by our data provider) so that  
PCA or more classical autoencoder approaches are also available. 
 The results show that the functional approach then performs as well as these classical benchmarks (which, however, would not be available on the original data 
 with variable time-to-maturity). 
 

A swaption is a financial contract allowing a client to enter into an interest rate swap with some strike $K$ at some future expiry date $U$, for some tenor length $T$.
A large body of literature deals with the swaption implied volatility as a function of the strike parameter. 

\begin{figure}[!htbp]
   \centering
   \includegraphics[width=.48\textwidth]{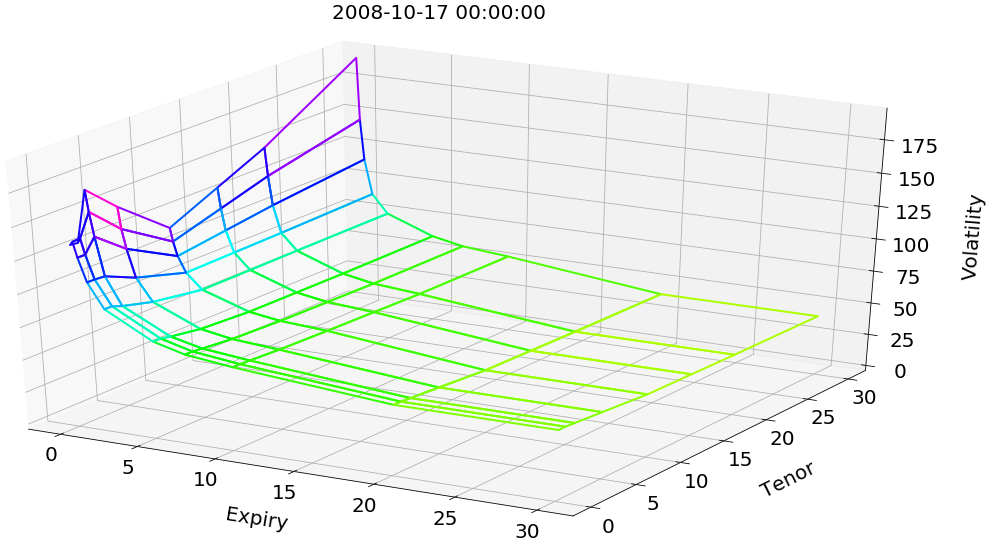}\quad
   \includegraphics[width=.48\textwidth]{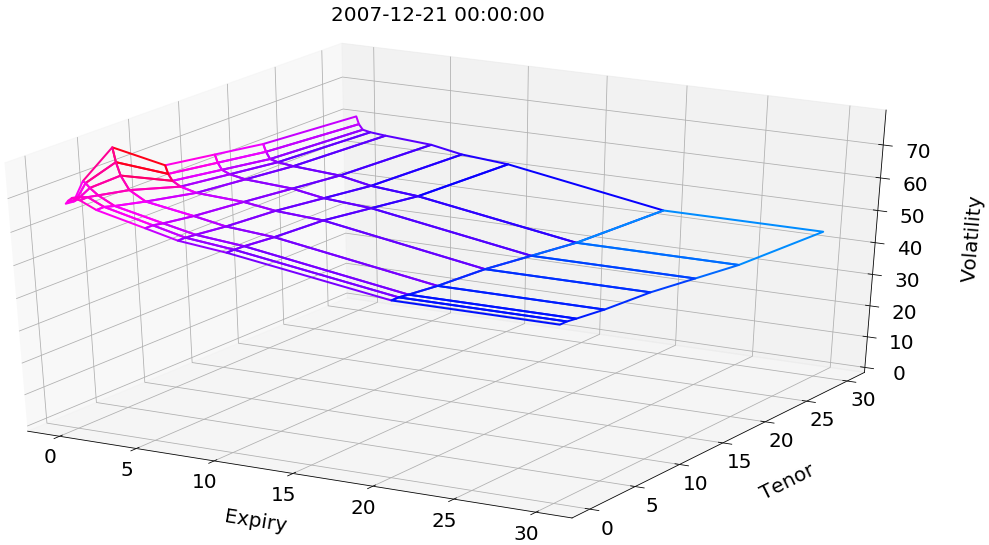}\\
   \includegraphics[width=.48\textwidth]{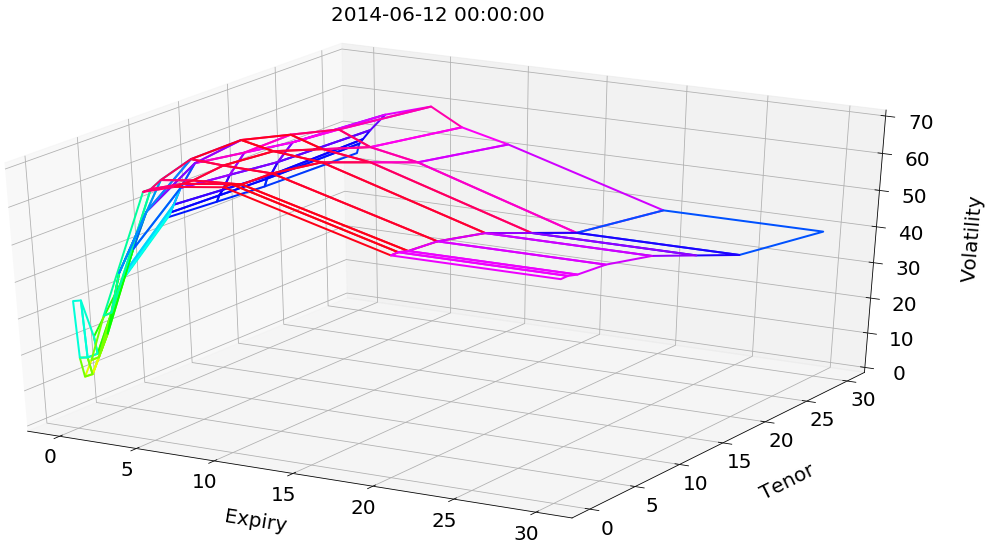}\quad
   \includegraphics[width=.48\textwidth]{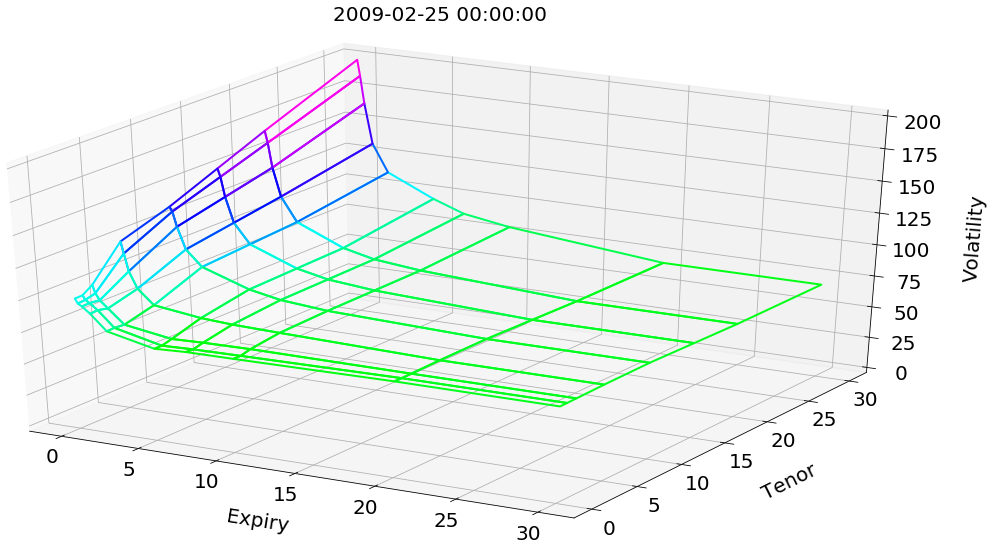}
   \caption{Different patterns of at-the-money swaption volatility surface.}
   \label{fig:sub1}
\end{figure}

By contrast, very few works are dealing with the
swaption implied volatility as a function of the expiry and tenor parameters  (see Figure \ref{fig:sub1}). 
One exception is \citeN{NBERw16549}, who,
based on a time series of swaption cubes, 
investigate how the conditional moments of the underlying swap rate distributions vary with expiry, tenor, and
calendar time.
One possible reason 
for this relative lack of literature
may be that swaption arbitrage pricing relationships are mainly known along the
strike direction. Across 
expiries and tenors, one only has ``statistical arbitrage'' relations,
reflecting the overlap between the cash flow streams of the underlying swaps.

In the following case study, we focus on at-the-money (ATM,
which are also the most liquid) swaption implied volatilities as a function of $U$ and  $T$.
The approach is model free in the sense that we do not formulate or use any hypothesis on the underlying forward swap rate processes.

Our study is conducted on a daily database of monocurrency (euro) ATM swaption normal\footnote{rather than Black--Scholes, because of the negative rates environment.} implied volatilities, covering 2400 business days corresponding to the period from 2007 to 2017.  The  training calibration and validation set $ \Omega $ covers the 2007 to 2014 sub-period (1900 first observation days of the data set), whereas the test set $\Omega'$ ranges from 2015 to 2017 (500 subsequent ones).
The data have been preprocessed by our provider so that all the ATM implied volatility surfaces are defined on a common rectangular grid of eighty $(U,T)$ nodes, without missing implied volatility values at any day or node, 
corresponding to the ten expiries (with M for month and Y for year)
$$U\in (1 M, 3M, 6M, 1Y, 2Y, 5Y, 7Y, 10Y, 20Y, 30Y)$$
and the eight tenors 
$$T\in (3M, 1Y, 2Y, 5Y, 10Y, 15Y, 20Y, 30Y).$$

For  testing our completion approach, we mask 90\% of the points in each surface of the test set $\Omega'$, only keeping the volatility points 
corresponding to the grid nodes $(U,T)$ in
\begin{eqnarray}\label{eq:unmask}\begin{aligned}& (1M,3M),(1M,10Y),(1M,30Y),(6M,2Y),\\&(6M,15Y),(5Y,1Y),(5Y,20Y),(10Y,5Y).\end{aligned}\end{eqnarray}
Such specification is in line with the reality of a market where the shortest expiries are the most liquidly traded ones (as well as the most volatile). Hence, our completion exercise corresponds to the intraday situation of a swaption trader facing mostly short expiry ATM implied volatility data, and left with the task of guessing the ``most likely values'' of the remaining implied volatilities. 

\subsection{Network Architectures} 

\begin{figure}[!htbp] 
  \centering
  \includegraphics[width=\linewidth]{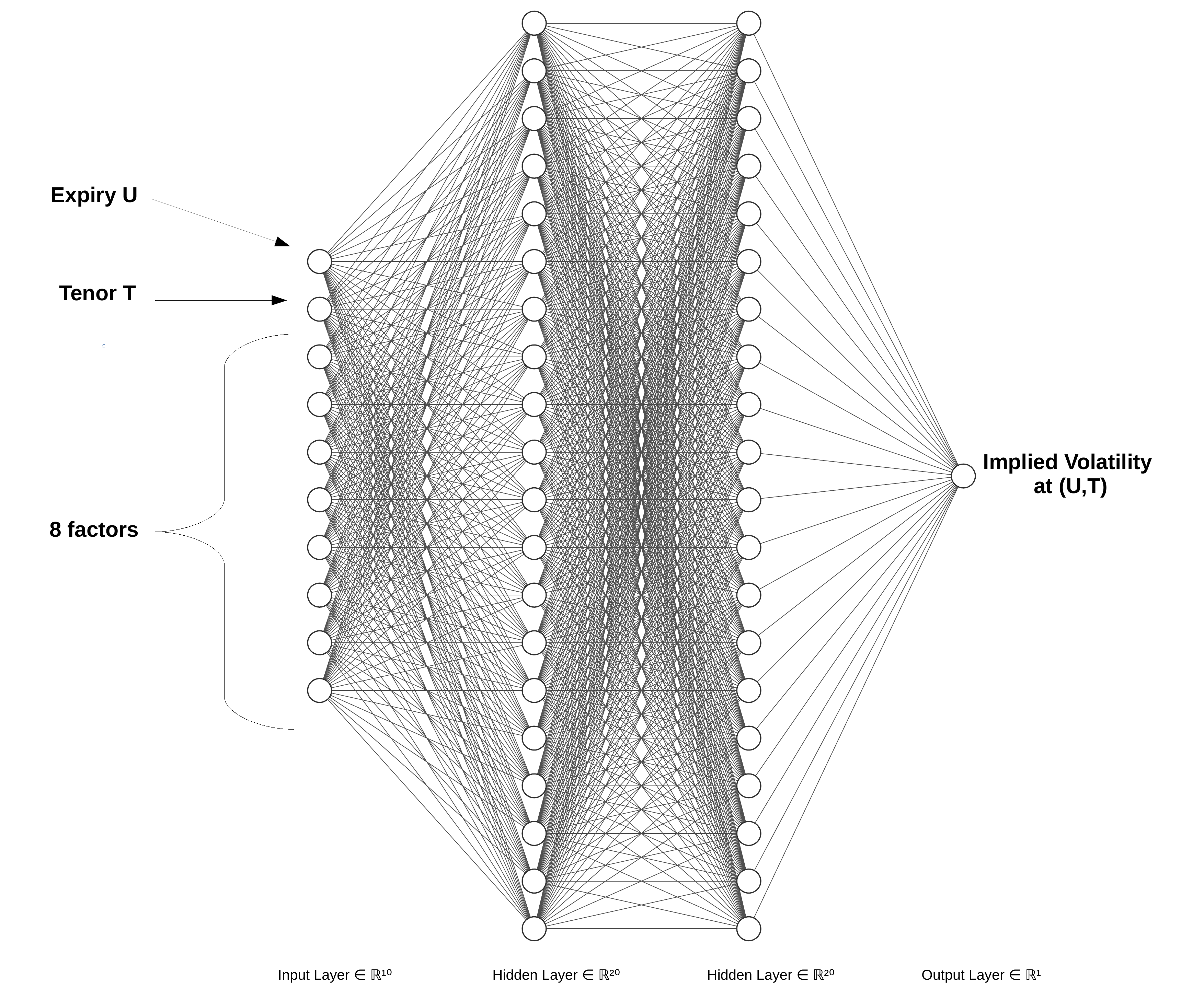}
  \caption{Network of the \functional approach used in
  the swaption
  case study (style FCNN of the NN-SVG software, cf. Figure \ref{fig:jiangNN}).}
  \label{fig:jiang} 
\end{figure}

The corresponding architecture of the \functional approach is then similar to the one used for equity derivatives in Section \ref{ss:ftuning},
except that the expiry $U$ and tenor $T$ are used as the localization inputs, and only 8 latent variables are used (instead of 15 previously): see Figure \ref{fig:jiang}.
Moreover, 
one can also incorporate the forward swap rates 
 as exogenous variables.
 These are the underlyings of the swaptions and they are structured similarly to the ATM implied volatilities of the latter, located by an expiry and a tenor.
For taking them into account, it suffices to add to the network
of Figure \ref{fig:jiang} an additional feature (input unit) containing the level of the  forward swap rate with  expiry $U$ and  tenor $T$. 
Hence,
the units for the
 expiry $U$ and the tenor $T$ 
indicate the common location of the corresponding 
 ATM 
volatilities and forward 
 swap 
rates.
 
 The convolutional autoencoders use feed-forward neural networks for the encoder and the decoder, with four hidden layers each:
one dense layer is applied on top of three convolutional layers for the encoder and, symmetrically, three deconvolutional layers are built on top of one dense layer.
The data set is reshaped as a $(10,8)$ tensor per day.  
The convolution layers are built with the respective kernels (used for specifying the localization of the weights) $(5,4)$, $(4,3)$, and $(3,3)$.
\begin{figure}[!htbp]
  \centering
  \includegraphics[width=\linewidth]{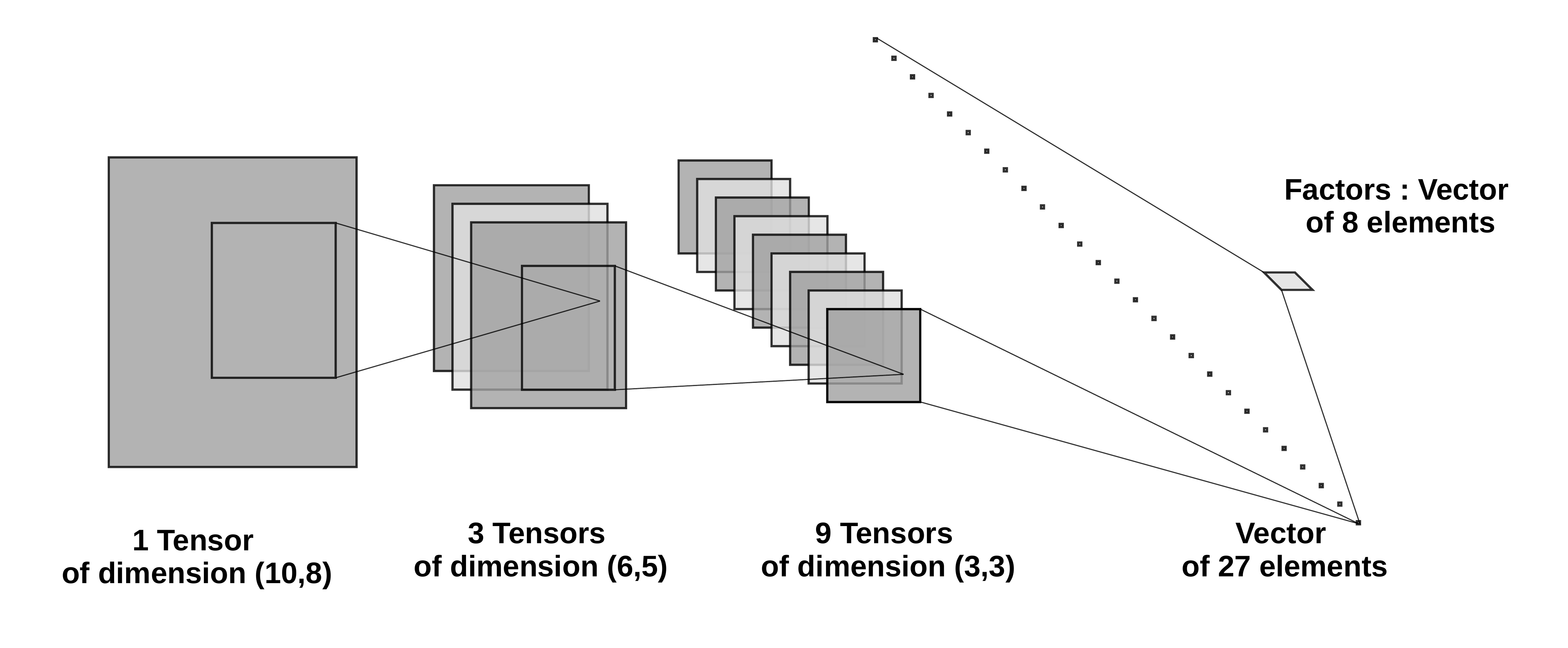} 
  \caption{Architecture of the convolutional encoder used in our ATM swaption case study. Graph produced using the style LeNet of the NN-SVG software: Each of the four layers is represented by a triangle; The inputs of each of the three convolutional layers are displayed as  collections of tensors; The ones of the last, dense layer are represented as a series of dots.}
  \label{fig:convEncoder} 
\end{figure}
Each convolution layer produces 3 channels (see Figure \ref{fig:convEncoder}) and, symmetrically, each deconvolution layer has in input 3 times more channels than in output.
Padding is set as VALID in order to reduce the size of the hidden units after each convolution layer.
As output of the three convolution layers, we have a hidden layer of 27 units, corresponding to 27 channels of size $(1,1)$.
A softplus (regularized ReLU) activation function is chosen after each convolution layer. This results in sparsity of the calibrated network (the compression stage sets very negative biases on the intermediate units that the neural network wants to ignore, cf.~\citeN{bengio2012practical}), as well as positivity and regularity of the ensuing implied volatility surface.
The dense layers between the factors and the (de)convolution layers are linear. Hence,
the convolution layers can be seen as a kernel that linearly separates the features.

Following a divide-and-conquer, sequential training strategy, we train the convolutional layers by pairs, from the most outer to the most inner ones, i.e.~the layers surrounding the latent variables (greedy layer-wise pre-training as per \citeN{hinton2006fast} and \citeN{bengio2007greedy}).
A final optimization fine-tunes the weights of all the layers together.
This also allows
exploiting any hierarchical structure of the data (cf.~\citeN{masci2011stacked}): The outer layers detect the greatest patterns, while inner layers detect the finest ones.

In the case of the fully connected networks that are used in the linear projection and in the \functional approaches,
we use the \citeN{Glorot} initialization rule for the weights,  with a centered normal distribution of standard deviation equal to $\sqrt{\frac{4}{n_{inputs} + n_{outputs}}}$. In the case of the convolutional layers we use a truncated normal distribution with 0.1 standard deviation. 
All biases are initialized to zero.

Each iteration leads to the computation of the loss gradient on the whole calibration data set.
Indeed, given the relatively small size of our data sets, full gradient evaluation is not an issue in practice. 
Moreover, mini-batch would require that each batch sample has approximately the same distribution, which is notoriously violated in the case of (non-stationary) financial time series. 

 Penalization is used at the compression stage for regularizing the calibrated parameters. More precisely,
ridge regularization is used for the kernel weights of the fully-connected layers of the convolutional and of the \functional approaches, with a penalization coefficient of $0.1$ intended to  balance the reconstruction loss and the penalization term at the minimum.

  \subsection{Numerical Results} 

Table \ref{tab:level} is a report on the errors of all our approaches (cf.~Section \ref{ss:casestud}).
  It is based on the absolute daily RMSEs (cf. \eqref{eq:etdun} and \eqref{eq:completionLoss}).
{\def\textbf#1{}
\footnotesize
{\hspace{5cm}
\begin{table}[!htbp]
\centering
 \resizebox{\textwidth}{!}{%
\begin{tabular}{|c|c|c|c|c|c|}
\hline
 & \begin{tabular}[c]{@{}c@{}}Standard\\ PCA \end{tabular} & \begin{tabular}[c]{@{}c@{}}Linear \\ projection\end{tabular} & \begin{tabular}[c]{@{}c@{}}Convolutional \\ autoencoder\end{tabular} &\begin{tabular}[c]{@{}c@{}}\Functional \\ approach\end{tabular}               & \begin{tabular}[c]{@{}c@{}}\Functional approach \\ with forward rate\end{tabular} \\ \hline\hline
\begin{tabular}[c]{@{}c@{}}Average compression   error\\ on                                        $\Omega$\end{tabular}      & 1.23  \textbf{(1.87\%)}                                                                       & 1.58  \textbf{(2.36\%)}                                                                         & 1.97   \textbf{(3.06\%)}                                                                      & 1.85  \textbf{(2.75\%)}                                                                               & 2.29  \textbf{(3.59\%)}                                                                       \\ \hline
\begin{tabular}[c]{@{}c@{}}Average compression   error \\on                                        $\Omega'$\end{tabular}     & 3.71  \textbf{(11.6\%)}                                                                      & 3.54  \textbf{(8.83\%)}                                                                       & 6.19  \textbf{(20.7\%)}                                                                      & 3.77  \textbf{(9.40\%)}                                                                              & 3.02  \textbf{(8.42\%)}                                                                       \\ \hline\hline
\begin{tabular}[c]{@{}c@{}}Worst compression   error \\on                                        $\Omega$ {[day] ([day]) }      \\       \end{tabular}  & \begin{tabular}[c]{@{}c@{}}4.15 \textbf{(7.87\%)}                                                                     \\ {[2008-12-03]} \\\textbf{{([                                        2014-06-10])}}                         \end{tabular}      & \begin{tabular}[c]{@{}c@{}} 3.98    \textbf{(6.92\%)}                                                                  \\ {[2008-12-09]} \\\textbf{{([                                        2014-06-10])}}                         \end{tabular}        & \begin{tabular}[c]{@{}c@{}}7.18    \textbf{(11.5\%)}                                                                  \\ {[2008-12-08]} \\\textbf{{([                                        2014-06-10])}}                           \end{tabular}        & \begin{tabular}[c]{@{}c@{}}8.32    \textbf{(15.1\%)}                                                                  \\ {[2008-10-09]} \\\textbf{{([                                        2014-06-06])}}                         \end{tabular} & \begin{tabular}[c]{@{}c@{}}6.93 \textbf{(12.2\%)}                                                                  \\ {[2008-10-10]} \\\textbf{{([                                        2014-06-13])}}                         \end{tabular}     \\ \hline
\begin{tabular}[c]{@{}c@{}}Worst compression   error \\on                                        $\Omega'$ {[day] ([day]) }     \\        \end{tabular} & \begin{tabular}[c]{@{}c@{}}5.76 \textbf{(23.5\%)}                                                                  \\ {[2016-04-28]} \\\textbf{{([                                        2016-04-28])}}                         \end{tabular}  & \begin{tabular}[c]{@{}c@{}}5.18 \textbf{(22.6\%)}                                                                  \\ {[2016-04-28]} \\\textbf{{([                                        2016-04-28])}}                           \end{tabular}    & \begin{tabular}[c]{@{}c@{}}12.0 \textbf{(55.1\%)}                                                                  \\ {[2015-07-07]} \\\textbf{{([                                        2016-04-28])}}                           \end{tabular}    & \begin{tabular}[c]{@{}c@{}}6.34 \textbf{(16.6\%)}                                                                  \\ {[2015-12-21} \\\textbf{{([                                        2014-10-09])}}                         \end{tabular} & \begin{tabular}[c]{@{}c@{}}5.16 \textbf{(16.4\%)}                                                                  \\ {[2015-12-18]} \\\textbf{{([                                        2014-07-23])}}                         \end{tabular}         \\ \hline\hline
\begin{tabular}[c]{@{}c@{}}Average completion  error\\  on $\Omega'$       \end{tabular}      & 6.19 \textbf{(16.9\%)}                                                                    & 4.07 \textbf{(10.4\%)}                                                                    & 5.03 \textbf{(13.5\%)}                                                                   & 6.41 \textbf{(15.5\%)}                                                                    & 5.19 \textbf{(13.8\%)}                                                                   \\ \hline
\begin{tabular}[c]{@{}c@{}}Worst completion   error \\on                                        $\Omega'$ {[day] ([day]) }    \\         \end{tabular}  & \begin{tabular}[c]{@{}c@{}}12.6 \textbf{(32.8\%)}                                                                  \\ {[2015-06-30]} \\\textbf{{([                                        2015-06-30])}}                         \end{tabular}  & \begin{tabular}[c]{@{}c@{}}6.50 \textbf{(19.2\%)}                                                                  \\ {[2015-07-10]} \\\textbf{{([                                        2016-04-28])}}                         \end{tabular}   & \begin{tabular}[c]{@{}c@{}}9.89 \textbf{(33.2\%)}                                                                  \\ {[2015-07-10]} \\\textbf{{([                                        2015-07-10])}}                         \end{tabular}      & \begin{tabular}[c]{@{}c@{}}12.8 \textbf{(29.3\%)}                                                                  \\ {[2015-03-09]} \\\textbf{{([                                        2014-08-12])}}                          \end{tabular} & \begin{tabular}[c]{@{}c@{}}9.09 \textbf{(25.0\%)}                                                                  \\ {[2016-01-14]} \\\textbf{{([                                        2014-08-11])}}                          \end{tabular}   \\ \hline\hline
\begin{tabular}[c]{@{}c@{}}Training time    in seconds \end{tabular}                   & $\emptyset$                                                         & 9    & 411 & 1287  & 276  \\ \hline
\end{tabular}} 
\caption{RMSEs in the sense of  \eqref{eq:etdun} and \eqref{eq:completionLoss}
}
\label{tab:level}
\end{table}
}
}

The last row of Table \ref{tab:level} displays the corresponding training times for all but the standard PCA approach, which involves no training and is in fact much faster than all the others (as it essentially reduces to the inversion of an $m\times m$ matrix, with $m=80$).
The dates in brackets in the tables identify the observations corresponding to the worst 
errors.

\begin{figure}[!htbp]
\centering
  \includegraphics[width=.7\linewidth]{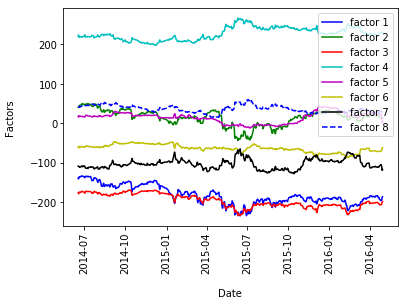}
  \caption{Time series of the factors obtained by encoding of the training observations under the linear projection approach.}
  \label{fig:factorsLevelVariation}
\end{figure}

At the completion stage, we take as initial factor values the volatility encoding of the previous day.
Figure \ref{fig:factorsLevelVariation} shows the stability  through calendar time of the codes obtained by the linear projection approach.
 
\begin{figure}[!htbp]
  \includegraphics[width=\linewidth]{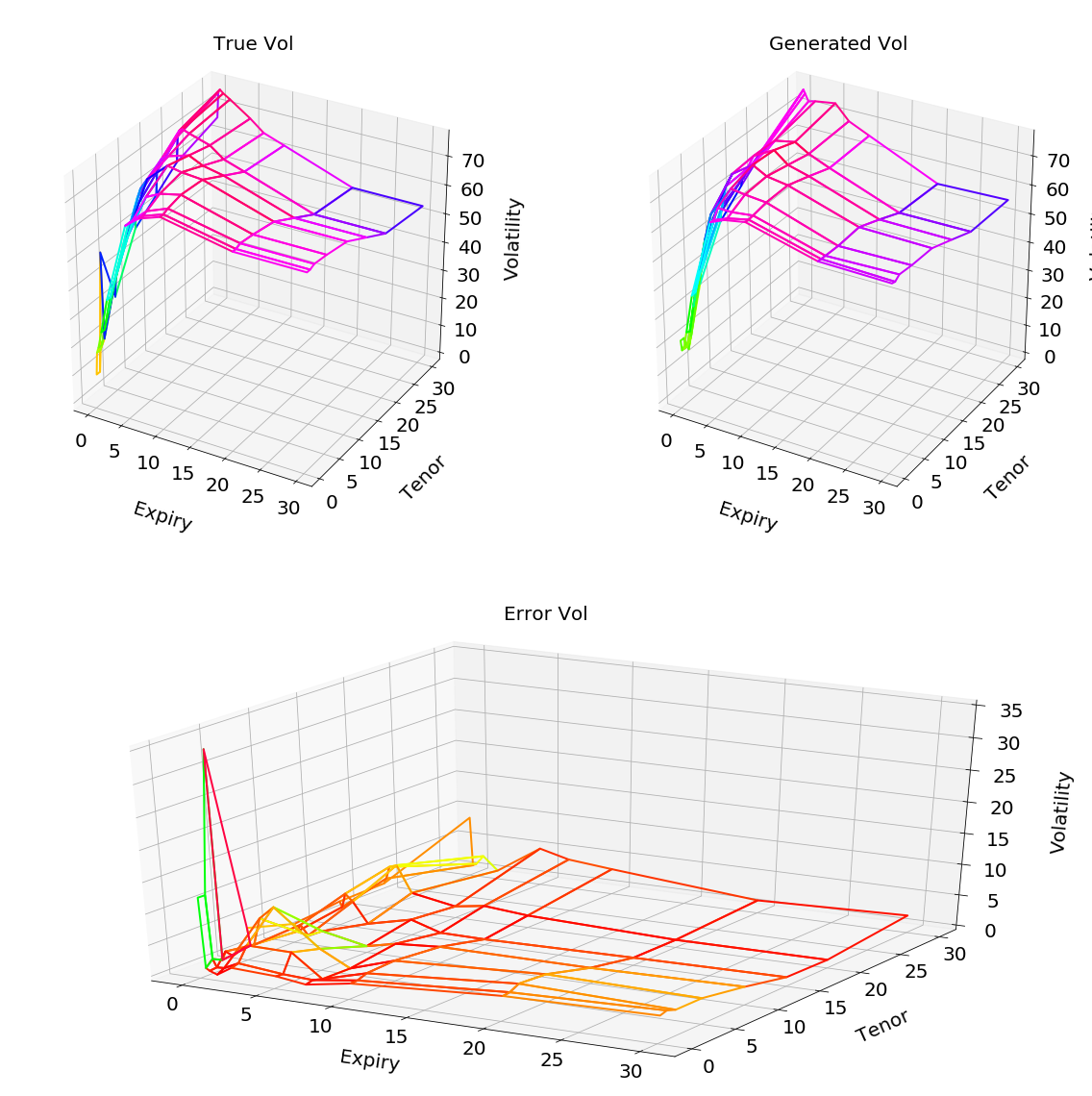}
  \caption{Linear projection approach: \textit{(Top left)} Original (full) tensor; \textit{(Top right)} Tensor $D_{\delta^{\star}}(c^{\star})$ completed based on the 8 points  of the latter given by \eqref{eq:unmask}; \textit{(Bottom)} Pointwise absolute error between the two, for the worst observation in $\Omega'$.}
  \label{fig:worstTestingCompressionLinear}
\end{figure}

As shown by Figure \ref{fig:worstTestingCompressionLinear}
in the case of the linear projection approach (but this is also true of the nonlinear approaches), 
the dominant errors are concentrated on the shortest expiries.
This is because the implied volatilities corresponding to these shortest expiries are the more volatile. Hence, their spatial dependence structure is less informative.
To recover these points better, one could think of providing extra information through exogenous variables, such as the level of the underlying forward swap rates. Under the \functional approach, this can easily be done in the way explained in Section \ref{ss:ftuning}. However, the last columns in Table
shows that this only has a minor positive impact.

The linear approaches are as accurate as the nonlinear ones and the convolutional approach is typically outperformed by at least the linear projection or the \functional approach.

\begin{figure}[!htbp]
  \includegraphics[width=\linewidth]{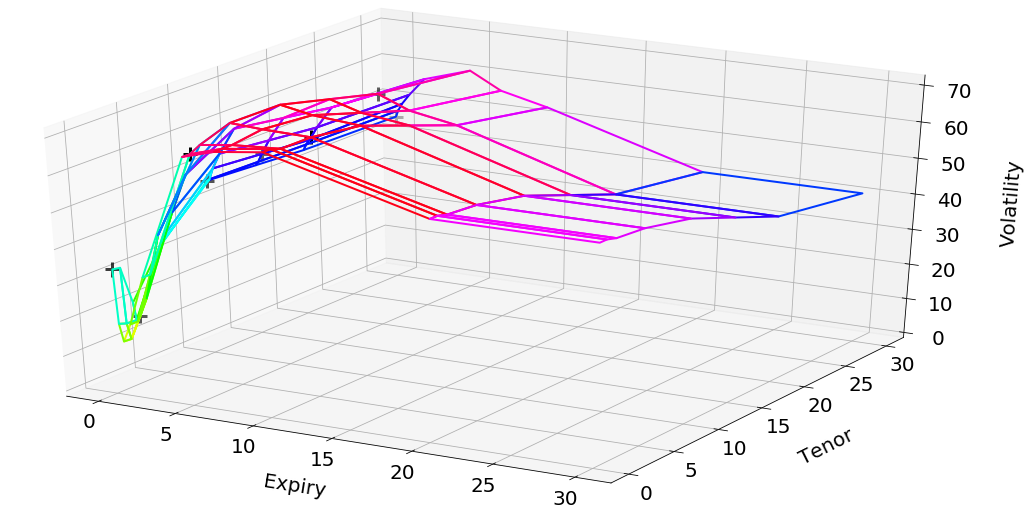}
  \caption{Complete tensor corresponding to the first observation in $\Omega'$. The black crosses designate the  ``\available points'', specified by \eqref{eq:unmask}, that are used in the completion exercise.} 
  \label{fig:originalSurface}   
\end{figure}
\begin{figure}[!htbp] 
  \includegraphics[width=\linewidth]{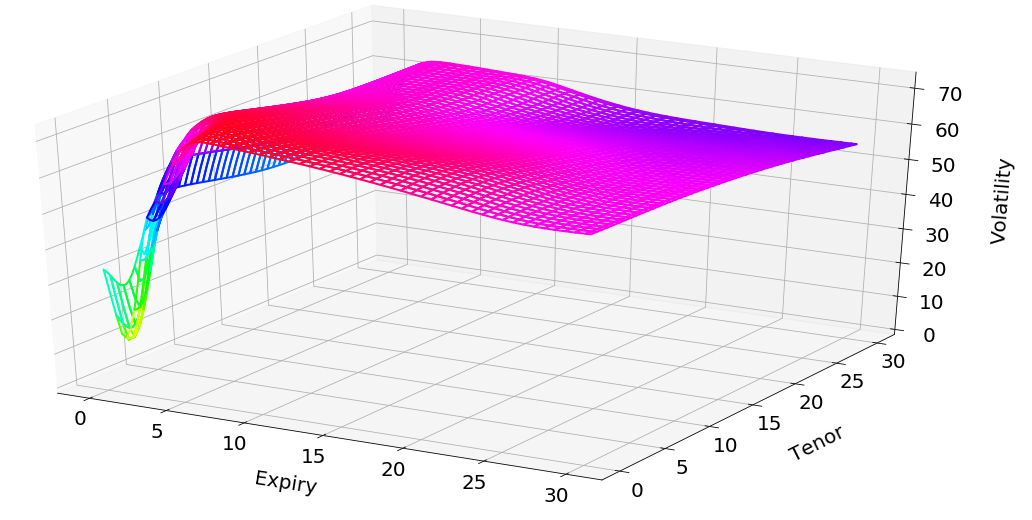}
  \caption{Surface with $10^4$ points obtained by the \functional approach applied to the first observation in $\Omega'$.}
  \label{fig:interpolatedSurface}
\end{figure}

%
Figure \ref{fig:interpolatedSurface} illustrates that the functional approach enables to interpolate smoothly the surface over an arbitrarily fine grid, in this case
$10^4$ points obtained by the corresponding interpolation of the tensor of Figure  \ref{fig:originalSurface}.

\section{Conclusions and Perspectives}\label{ss:concl}

We have devised a generic neural network based curve or surface (or more general tensor) compression/completion methodology, 
for which we
propose two concrete specifications: the   \functional approach,   amenable to the treatment of unstructured data with varying grid nodes (as natively the case in most financial nowcasting applications), and 
a \convolutional autoencoder approach, including PCA or PCA-like projections as linear special cases, applicable  in the special case of a constant grid (natively or possibly after some preprocessing). 
The compression stage also allows for outlier detection and correction by generating surfaces or curves in line with training samples. 

The analysis of the corresponding reconstruction errors suggests that linear methods are sufficient to compress structured tensors, corresponding to a constant grid of nodes, into few factors coefficients. The completion stage
allows recovering with success
 about 90\% values of the data, starting from about 10\% of known values.
But  the \functional approach is the only one that is able to directly deal (without preprocessing) with the most common situation of unstructured tensors. 
The only alternative is then naive interpolation benchmarks that do not exploit the data set, and which the functional approach is shown to outperform in our equity derivative case study.

All approaches suffer from non-stationarities occurring during extreme events or change of market regimes.
This can be seen as an advantage with respect to anomaly detection. For other purposes, it would plead in favor of further modeling of the factor dynamics, whether this relies on times series machine learning or Markov chain Monte Carlo (filtering) techniques.
More generally, it would be interesting to extend this study in several directions, such as
the introduction of backtesting hedging criteria (cf. \citeN{garcia2000pricing}), scenario simulation in a context of variational networks (see \citeN{VAEDisentangle}), application of the  method to the whole swaption volatility cube, strike dimension included (cf.~\citeN{NBERw16549}), or specification of dynamics on the factors (for instance by Kalman filters).

\bibliographystyle{chicago}

\end{document}